\documentclass[12pt,thmsa]{article}
\usepackage{amsfonts}
\usepackage{amsmath}
\usepackage{sw20lart}

\input tcilatex
\begin{document}

\author{Nick Laskin\thanks{\textit{E-mail address}: nlaskin@rocketmail.com}}
\title{\textbf{Some Applications of the Fractional Poisson Probability
Distribution}}
\date{TopQuark Inc.\\
Toronto, ON, M6P 2P2\\
Canada}
\maketitle

\begin{abstract}
Physical and mathematical applications of fractional Poisson probability
distribution have been presented. As a physical application, a new family of
quantum coherent states has been introduced and studied. As mathematical
applications, we have discovered and developed the fractional generalization
of Bell polynomials, Bell numbers, and Stirling numbers. Appearance of
fractional Bell polynomials is natural if one evaluates the diagonal matrix
element of the evolution operator in the basis of newly introduced quantum
coherent states. Fractional Stirling numbers of the second kind have been
applied to evaluate skewness and kurtosis of the fractional Poisson
probability distribution function. A new representation of Bernoulli numbers
in terms of fractional Stirling numbers of the second kind has been
obtained. A representation of Schl\"{a}fli polynomials in terms of
fractional Stirling numbers of the second kind has been found. A new
representations of Mittag-Leffler function involving fractional Bell
polynomials and fractional Stirling numbers of the second kind have been
discovered. Fractional Stirling numbers of the first kind have been
introduced and studied. Two new polynomial sequences associated with
fractional Poisson probability distribution have been launched and explored.
The relationship between new polynomials and the orthogonal Charlier
polynomials has also been investigated.

In the limit case when the fractional Poisson probability distribution
becomes the Poisson probability distribution, all of the above listed
developments and implementations turn into the well-known results of quantum
optics, the theory of combinatorial numbers and the theory of orthogonal
polynomials of discrete variable.


\end{abstract}

\tableofcontents

\section{Introduction}

In the past decade it has been realized that modeling complex quantum and
classical physics phenomena requires the implementation of long-range space
and long-memory stochastic processes. The mathematical model to capture
impact of long-range phenomena at a quantum level is the \textit{L\'{e}vy
path integral approach} invented and studied in Refs.\cite{Laskin1}-\cite%
{Laskin6}. The L\'{e}vy path integral approach generalizes the path integral
formulation of quantum mechanics developed in 1948 by Feynman \cite{Feynman1}%
, \cite{Feynman2}. The generalization results in \textit{fractional quantum
mechanics} \cite{Laskin1}-\cite{Laskin3}. One of the \ fundamental equations
of fractional quantum mechanics is the \textit{fractional Schr\"{o}dinger
equation} discovered in \cite{Laskin1}-\cite{Laskin4}, \cite{Laskin6}. The
fractional Schr\"{o}dinger equation is a new non-Gaussian physical paradigm,
based on deep relationships between the structure of fundamental physics
equations and fractal dimensions of \textquotedblleft
underlying\textquotedblright\ quantum paths.

To study a long-memory impact on the counting process, the \textit{%
fractional Poisson process} has been invented and developed for the first
time by Laskin in Ref.\cite{Laskin7}. The fractional Poisson probability
distribution captures the long-memory effect which results in the
non-exponential waiting time probability distribution function \cite{Laskin7}%
, empirically observed in complex quantum and classical systems. The quantum
system example is the fluorescence intermittency of single CdSe quantum
dots, that is, the fluorescence emission of single nanocrystals exhibits
intermittent behavior, namely, a sequence of \textquotedblright light
on\textquotedblright\ and \textquotedblright light off\textquotedblright\
states departing from Poisson statistics. In fact, the waiting time
distribution in both states is non-exponential \cite{Kuno}. As examples of
classical systems let's mention the distribution of waiting times between
two consecutive transactions in financial markets \cite{Sabatelli} and
another, which comes from network communication systems, where the duration
of network sessions or connections exhibits non-exponential behavior \cite%
{Willinger}.

The non-exponential waiting time distribution function has been obtained for
the first time in Ref.\cite{Zaslavsky}, based on the fractional
generalization of the Poisson exponential waiting time distribution.

The fractional Poisson process is a natural generalization of the well known
Poisson process. A simple analytical formula for the fractional Poisson
probability distribution function has been obtained for the first time in
Ref.\cite{Laskin7} based on the fractional generalization of the
Kolmogorov-Feller equation introduced in Ref.\cite{Zaslavsky}. It was shown
by Laskin in \cite{Laskin7} that the non-exponential waiting time
distribution function of fractional Poisson process obtained in \cite%
{Laskin7} is identical to the one found in \cite{Zaslavsky}.

In comparison to standard Poisson distribution, the probability distribution
function of the fractional Poisson process \cite{Laskin7} has an additional
parameter $\mu $, $0<\mu \leq 1$. In the limit case $\mu =1$ the fractional
Poisson process becomes the standard Poisson process and all our findings
are transformed into the well-known results related to the standard Poisson
probability distribution.

Now we present quantum physics and mathematical applications of the \textit{%
fractional Poisson probability distribution }\cite{Laskin8}. This paper is
an extended version of articles \cite{Laskin8}, \cite{Laskin9}.

The quantum physics application is an introduction of a new family of
quantum coherent states. The motivation to introduce and explore these
coherent states is the observation that the squared modulus $|<n|\varsigma
>|^{2}$ of projection of the newly invented coherent state $|\varsigma >$
onto the eigenstate of the photon number operator $|n>$ gives us the
fractional Poisson probability $P_{\mu }(n)$ that $n$ photons will be found
in coherent state $|\varsigma >.$ Following Klauder's framework to qualify
quantum states as generalized coherent states \cite{Klauder1}, we prove that
our quantum coherent states $|\varsigma >$, (i) are parametrized
continuously and normalized; (ii) admit a resolution of unity with positive
weight function; (iii) provide temporal stability, that is, the time
evolution of coherent states remains within the family of coherent states.
We have defined the inner product of two vectors in terms of their coherent
state $|\varsigma >$ representation and introduced functional Hilbert space.

Mathematical applications are related to number theory and theory of
polynomials of discrete variable. Bell polynomials, Bell numbers \cite{Bell}
and Stirling numbers \cite{Stirling} - \cite{Charalambides} have been
generalized based on the fractional Poisson probability distribution. In
other words, based on fractional Poisson probability distribution, we
introduce new fractional Bell polynomials, new fractional Bell numbers and
new fractional Stirling numbers in the same fashion as the well-known Bell
polynomials, Bell numbers and Stirling numbers can be introduced based on
the famous Poisson probability distribution. Appearance of fractional Bell
polynomials is natural if one evaluates the diagonal matrix element of the
quantum evolution operator in the basis of newly introduced quantum coherent
states. The appearance of fractional Bell numbers manifests itself in the
fractional generalization of the celebrated Dobi\'{n}ski formula \cite%
{Dobinski}, \cite{Rota} for the generating function of the Bell numbers.

Fractional Stirling numbers of the second kind have been applied to evaluate
skewness and kurtosis of the fractional Poisson probability distribution.

Fractional Stirling numbers of the first kind have also been introduced and
studied.

A representation of Schl\"{a}fli polynomials in terms of fractional Stirling
numbers of the second kind has been found. The integral relationship between
the Schl\"{a}fli polynomials and fractional Bell polynomials has been
obtained. A new representation of the Mittag-Leffler function involving
fractional Bell polynomials and fractional Stirling numbers of the second
kind has been discovered.

New polynomials of discrete variable associated with fractional Poisson
probability distribution have been launched and explored in the
multiplicative renormalization \cite{Kubo} framework.

In the limit case when $\mu =1$ and the fractional Poisson probability
distribution becomes the standard Poisson probability distribution, all
above listed new developments and findings turn into the well-known results
of the quantum coherent states theory \cite{Klauder}-\cite{Wolf}, the theory
of combinatorial numbers \cite{Charalambides1}, \cite{Charalambides} and the
theory of orthogonal polynomilas of discrete variable \cite{Chihara}.

The paper is organized as follows.

Basic definitions of the fractional Poisson random process are briefly
reviewed in Sec.2, where Table 1 has been presented to compare the formulas
related to the fractional Poisson probability distribution \cite{Laskin7} to
those of the well-known ones, related to the standard Poisson probability
distribution. In Sec.3 we introduce and study new quantum coherent states
and their applications. Fractional generalizations of Bell polynomials, Bell
numbers and Stirling numbers have been introduced and developed in Sec.4.
New equations for the generating functions of fractional Bell polynomials,
fractional Bell numbers and fractional Stirling numbers have been obtained
and elaborated. The relationship between Bernoulli numbers and fractional
Stirling numbers of the second kind has been found. A new representation of
the Schl\"{a}fli polynomials in terms of fractional Stirling numbers of the
second kind has been found. The integral relationship between the Schl\"{a}%
fli polynomials and fractional Bell polynomials has been found. A new
representations of the Mittag-Leffler function involving fractional Bell
polynomials and fractional Stirling numbers of the second kind have been
discovered and explored. Fractional Stirling numbers of the first kind have
also been introduced and studied in Sec.4.

Moments and central moments of the fractional Poisson probability
distribution have been studied in Sec.5. The central moment of $m$-order has
been obtained in terms of fractional Stirling numbers of the second kind.
Variance, skewness and kurtosis of the fractional Poisson probability
distribution function have been presented in terms of the central moments of 
$m$-order.

A new class of polynomials of discrete variable associated with fractional
Poisson probability distribution have been launched and explored in Sec.6.
It has been observed that in the limit case when $\mu =1$, these new
polynomials become the well-known Charlier orthogonal polynomials \cite%
{Chihara}.

Table 2 summarizes key fundamental equations of the coherent states theory
for new coherent states $|\varsigma >$ vs those for standard coherent states 
$|z>$. Table 3 displays the moment generating functions of four fractional
compound Poisson processes. Table 4 presents a few fractional Stirling
numbers of the second kind. Table 5 presents polynomials, numbers, moments
and generating functions attributed to the fractional Poisson probability
distribution vs the standard Poisson probability distribution. Table 6
presents a few fractional Stirling numbers of the first kind. Table 7
compares fundamentals for newly introduced polynomials $L_{n}(x;\lambda
_{\mu })$\ vs the Charlier orthogonal polynomials $C_{n}(x;\lambda )$.

\section{Fundamentals of the fractional Poisson probability distribution}

The fractional Poisson process has originally been introduced and developed
by Laskin \cite{Laskin7} as the counting process with probability $P_{\mu
}(n,t)$ of arriving $n$ items ($n=0,1,2,...$) by time $t$. Probability $%
P_{\mu }(n,t)$ is governed by the system of fractional
differential-difference equations

\begin{equation}
_{0}D_{t}^{\mu }P_{\mu }(n,t)=\nu \left( P_{\mu }(n-1,t)-P_{\mu
}(n,t)\right) ,\quad n\geq 1,  \label{eq1}
\end{equation}

and

\begin{equation}
_0D_t^\mu P_\mu (0,t)=-\nu P_\mu (0,t)+\frac{t^{-\mu }}{\Gamma (1-\mu )}%
,\qquad 0<\mu \leq 1,  \label{eq1a}
\end{equation}

with normalization condition

\begin{equation}
\sum\limits_{n=0}^{\infty }P_{\mu }(n,t)=1.  \label{eq2}
\end{equation}

Here, $_{0}D_{t}^{\mu }$ is the operator of time derivative of fractional
order $\mu $ defined as the Riemann-Liouville integral\footnote{%
The basic formulas on fractional calculus can be found in Refs. \cite{Oldham}
- \cite{Miller}.},

\begin{equation*}
_{0}D_{t}^{\mu }\mathrm{f}(t)=\frac{1}{\Gamma (1-\mu )}\frac{d}{dt}%
\int\limits_{0}^{t}\frac{d\tau \mathrm{f}(\tau )}{(t-\tau )^{\mu }},\qquad
0<\mu \leq 1,
\end{equation*}

where $\mu $ is the fractality parameter, gamma function $\Gamma (\mu )$ has
the familiar representation $\Gamma (\mu )=\int\limits_{0}^{\infty
}dte^{-t}t^{\mu -1}$, $\mathrm{Re}\mu >0$, and parameter $\nu $ has physical
dimension $[\nu ]=\sec ^{-\mu }$. The system introduced by Eqs.(\ref{eq1})
and (\ref{eq1a}) has the initial condition $P_{\mu }(n,t=0)=\delta _{n,0}$.
One can consider the fractional differential-difference system of equations (%
\ref{eq1}) and (\ref{eq1a}) as a generalization of the
differential-difference equations which define the well-known Poisson
process (see, for instance, Eqs.(6) and (7) in Ref.\cite{Laskin7}).

To solve the system of equations (\ref{eq1}) and (\ref{eq1a}) it is
convenient to use the method of the generating function. We introduce the
generating function $G_{\mu }(s,t)$

\begin{equation}
G_{\mu }(s,t)=\sum\limits_{n=0}^{\infty }s^{n}P_{\mu }(n,t).  \label{eq3}
\end{equation}

Hence, to obtain $P_{\mu }(n,t)$ we have to calculate

\begin{equation}
P_{\mu }(n,t)=\frac{1}{n!}\frac{\partial ^{n}G_{\mu }(s,t)}{\partial s^{n}}%
|_{s=0}.  \label{eq3.1}
\end{equation}

Then, by multiplying Eqs.(\ref{eq1}) and (\ref{eq1a}) by $s^{n}$, summing
over $n$, we obtain the following fractional differential equation for the
generating function $G_{\mu }(s,t)$

\begin{equation}
_{0}D_{t}^{\mu }G_{\mu }(s,t)=\nu \left( \sum\limits_{n=0}^{\infty
}s^{n}P_{\mu }(n-1,t)-\sum\limits_{n=0}^{\infty }s^{n}P_{\mu }(n,t)\right) =
\label{eq4}
\end{equation}

\begin{equation*}
\nu (s-1)G_{\mu }(s,t)+\frac{t^{-\mu }}{\Gamma (1-\mu )}.
\end{equation*}

The solution of this fractional differential equation has been found in \cite%
{Laskin7}

\begin{equation}
G_\mu (s,t)=E_\mu (\nu t^\mu (s-1)),  \label{eq5}
\end{equation}

where $E_{\mu }(z)$ is the Mittag-Leffler function\footnote{%
At $\mu =1$ the function $E_{\mu }(z)$ turns into $\exp (z)$.} given by its
power series \cite{ML}, \cite{Erdelyi}

\begin{equation}
E_\mu (z)=\sum\limits_{m=0}^\infty \frac{z^m}{\Gamma (\mu m+1)}.  \label{eq6}
\end{equation}

It follows from Eqs.(\ref{eq3.1}), (\ref{eq5}) and (\ref{eq6}) that

\begin{equation}
P_{\mu }(n,t)=\frac{(\nu t^{\mu })^{n}}{n!}\sum\limits_{k=0}^{\infty }\frac{%
(k+n)!}{k!}\frac{(-\nu t^{\mu })^{k}}{\Gamma (\mu (k+n)+1)},\qquad 0<\mu
\leq 1.  \label{eq7}
\end{equation}

This is the \textit{fractional Poisson probability distribution} obtained
for the first time by Laskin in \cite{Laskin7}. It gives us the probability
that in the time interval $[0,t]$ we observe $n$ counting events. When $\mu
=1$, $P_{\mu }(n,t)$ is transformed to the standard Poisson probability
distribution function (see Eq.(14) in Ref.\cite{Laskin7}). Thus, Eq.(\ref%
{eq7}) can be considered as a fractional generalization of the well-known
Poisson probability distribution function. The presence of an additional
parameter $\mu $ brings new features in comparison with the standard Poisson
probability distribution.

On a final note, the probability distribution of the fractional Poisson
process $P_{\mu }(n,t)$ can be represented in terms of the Mittag-Leffler
function $E_{\mu }(z)$ in the following compact way \cite{Laskin7},

\begin{equation}
P_{\mu }(n,t)=\frac{(-z)^{n}}{n!}\frac{d^{n}}{dz^{n}}E_{\mu }(z)|_{z=-\nu
t^{\mu }},  \label{eq8}
\end{equation}

\begin{equation}
P_{\mu }(n=0,t)=E_{\mu }(-\nu t^{\mu }).  \label{eq9}
\end{equation}%
At $\mu =1$ Eqs.(\ref{eq8}) and (\ref{eq9}) are transformed into the well
known equations for the famous Poisson probability distribution $P(n,t)$
with the substitution $\nu \rightarrow \overline{\nu }$, where $\overline{%
\nu }$ is the rate of arrivals of the Poisson process with physical
dimension $\overline{\nu }=\sec ^{-1}$,

\begin{equation}
P_{\mu }(n,t)|_{\mu =1}=P(n,t)=\frac{(\overline{\nu }t)^{n}}{n!}e^{-%
\overline{\nu }t},  \label{eq9.1}
\end{equation}

\begin{equation}
P(n=0,t)=e^{-\overline{\nu }t}.
\end{equation}

Table 1 compares equations attributed to the fractional Poisson process with
those belonging to the well-known standard Poisson process. Table 1 presents
two sets of equations for the probability distribution function $P(n,t)$ of $%
n$ events having arrived by time $t$, the probability $P(0,t)$ of having
nothing arrived by time $t,$ mean $\overline{n}$, variance $\sigma ^{2}$,
generating function $G(s,t)$ for the probability distribution function,
moment generating function $H(s,t)$, and the waiting-time probability
distribution function $\psi (\tau )$.

\begin{tabular}{|c|c|c|}
\hline
& fractional Poisson ($0<\mu \leq 1)$ & Poisson ($\mu =1)$ \\ \hline
$P(n,t)$ & $\frac{(\nu t^{\mu })^{n}}{n!}\sum\limits_{k=0}^{\infty }\frac{%
(k+n)!}{k!}\frac{(-\nu t^{\mu })^{k}}{\Gamma (\mu (k+n)+1)}$ & $\frac{(%
\overline{\nu }t)^{n}}{n!}\exp (-\overline{\nu }t)$ \\ \hline
$P(n,t)$ & $\frac{(-z)^{n}}{n!}\frac{d^{n}}{dz^{n}}E_{\mu }(z)|_{z=-\nu
t^{\mu }}$ & $\frac{(-z)^{n}}{n!}\frac{d^{n}}{dz^{n}}\exp (z)|_{z=-\overline{%
\nu }t^{\mu }}$ \\ \hline
$P(0,t)$ & $E_{\mu }(-\nu t^{\mu })$ & $\exp (-\overline{\nu }t)$ \\ \hline
$\overline{n}$ & $\frac{\nu t^{\mu }}{\Gamma (\mu +1)}$ & $\overline{\nu }t$
\\ \hline
$\sigma ^{2}$ & $\frac{\nu t^{\mu }}{\Gamma (\mu +1)}+\left( \frac{\nu
t^{\mu }}{\Gamma (\mu +1)}\right) ^{2}\left\{ \frac{\mu B(\mu ,\frac{1}{2})}{%
2^{2\mu -1}}-1\right\} $ & $\overline{\nu }t$ \\ \hline
$G(s,t)$ & $E_{\mu }(\nu t^{\mu }(s-1))$ & $\exp \{\overline{\nu }t(s-1)\}$
\\ \hline
$H(s,t)$ & $E_{\mu }(\nu t^{\mu }(e^{-s}-1))$ & $\exp \{\overline{\nu }%
t(e^{-s}-1)\}$ \\ \hline
$\psi (\tau )$ & $\nu \tau ^{\mu -1}E_{\mu ,\mu }(-\nu \tau ^{\mu })$ & $%
\overline{\nu }e^{-\overline{\nu }\tau }$ \\ \hline
\end{tabular}

Table 1.\textit{\ Fractional Poisson process vs the Poisson process}%
\footnote{%
All definitions and equations related to the fractional Poisson process are
taken from \cite{Laskin7}.}\textit{.}

\section{New family of coherent states}

The quantum mechanical states first introduced by Schr\"{o}dinger \cite%
{Schrodinger} to study the quantum harmonic oscillator are now well-known as
the coherent states. Coherent states provide an important theoretical
paradigm to study electromagnetic field coherence and quantum optics
phenomena \cite{Klauder}, \cite{Glauber}.

The standard coherent states are defined for all complex numbers $z\in C$, by

\begin{equation}
|z>=e^{(za^{+}-z^{\ast }a)}|0>=e^{-\frac{1}{2}|z|^{2}}\sum\limits_{n=0}^{%
\infty }\frac{z^{n}}{\sqrt{n!}}|n>,  \label{eq10}
\end{equation}

where $a^{+}$ and $a$ are photon field creation and annihilation operators
that satisfy the Bose-Einstein commutation relation $[a,a^{+}]=aa^{+}-a^{+}a=%
\QTR{sl}{1}$, and the orthonormal vector $|n>=\frac{1}{\sqrt{n!}}%
(a^{+})^{n}|0>$ is an eigenvector of the photon number operator $N=a^{+}a$, $%
N|n>=n|n>$, $<n|n^{\prime }>=\delta _{n,n^{\prime }}$. The action of the
operators $a^{+}$ and $\ a$ act on the state $|n>$ reads

{}%
\begin{equation*}
a^{+}|n>=\sqrt{n+1}|n+1>\qquad \text{and}\qquad a|n>=\sqrt{n}|n-1>.
\end{equation*}

The projection of coherent state $|z>$ onto state $|n>$ is

\begin{equation}
<n|z>=\frac{z^{n}}{\sqrt{n!}}e^{-\frac{1}{2}|z|^{2}}.  \label{eq11}
\end{equation}

Then the squared modulus of $<n|z>$ gives us probability $P(n)$ that $n$
photons will be found in the coherent state $|z>$. Thus, we come to the
well-know result for probability $P(n)$

\begin{equation}
P(n)=|<n|z>|^{2}=\frac{|z|^{2n}}{n!}e^{-|z|^{2}},  \label{eq12}
\end{equation}

which is recognized as a Poisson probability distribution with a mean value $%
|z|^{2}.$ The value $|z|^{2}$ is in fact the mean number of photons when the
state is a coherent state $|z>$

\begin{equation}
|z|^{2}=\sum\limits_{n=0}^{\infty }nP(n)=<z|a^{+}a|z>.  \label{eq13}
\end{equation}

We introduce a new family of coherent states $|\varsigma >$

\begin{equation}
|\varsigma >=\sum\limits_{n=0}^{\infty }\frac{(\sqrt{\mu }\varsigma ^{\mu
})^{n}}{\sqrt{n!}}(E_{\mu }^{(n)}(-\mu |\varsigma |^{2\mu }))^{1/2}|n>,
\label{eq14}
\end{equation}

and adjoint states $<\varsigma |$

\begin{equation}
<\varsigma |=\sum\limits_{n=0}^{\infty }<n|\frac{(\sqrt{\mu }\varsigma
^{\ast \mu })^{n}}{\sqrt{n!}}(E_{\mu }^{(n)}(-\mu |\varsigma |^{2\mu
}))^{1/2},  \label{eq15}
\end{equation}

where 
\begin{equation}
E_\mu ^{(n)}(-\mu |\varsigma |^{2\mu })=\frac{d^n}{dz^n}E_\mu (z)|_{z=-\mu
|\varsigma |^{2\mu }}  \label{eq16}
\end{equation}

and $E_\mu (z)$ is the Mittag-Leffler function defined by Eq.(\ref{eq6}),
complex number $\varsigma $ stands for labelling the new coherent states,
and the orthonormal vectors $|n>$ are the same as for Eq.(\ref{eq10}).

To motivate the introduction of new coherent states $|\varsigma >$ we
calculate the projection of coherent state $|\varsigma >$ onto state $|n>,$

\begin{equation*}
<n|\varsigma >=\frac{(\sqrt{\mu }\varsigma ^\mu )^n}{\sqrt{n!}}\left( E_\mu
^{(n)}(-\mu |\varsigma |^{2\mu })\right) ^{1/2},
\end{equation*}

then the squared modulus of $<n|\varsigma >$ gives us the probability $%
P_{\mu }(n)$ that $n$ photons will be found in the quantum coherent state $%
|\varsigma >$. Thus, we come to the fractional Poisson probability
distribution of photon numbers $P_{\mu }(n)$

\begin{equation}
P_{\mu }(n)=|<n|\varsigma >|^{2}=\frac{(\mu |\varsigma |^{2\mu })^{n}}{n!}%
\left( E_{\mu }^{(n)}(-\mu |\varsigma |^{2\mu })\right) ,  \label{eq17}
\end{equation}

with mean value ($\mu |\varsigma |^{2\mu })/\Gamma (\mu +1).$ The value ($%
\mu |\varsigma |^{2\mu })/\Gamma (\mu +1)$ is in fact the mean number of
photons when the quantum state is the coherent state $|\varsigma >$

\begin{equation*}
(\mu |\varsigma |^{2\mu })/\Gamma (\mu +1)=\sum\limits_{n=0}^\infty nP_\mu
(n)=<\varsigma |a^{+}a|\varsigma >.
\end{equation*}

It is easy to see that when $\mu =1$ we have $E_{1}(z)=\exp (z)$ and $%
E_{1}^{(n)}(-z)=\exp (-z)$. Hence, with substitution $\varsigma \rightarrow
z $ and $\mu =1$ Eq.(\ref{eq14}) turns into Eq.(\ref{eq10}), and Eq.(\ref%
{eq17}) leads to Eq.(\ref{eq11}). In other words, the new coherent states
defined by Eq.(\ref{eq14}) generalize the standard coherent states Eq.(\ref%
{eq10}), and this generalization has been implemented with the help of the
fractional Poisson probability distribution. The new family of coherent
states Eq.(\ref{eq14}) has been designed here to study physical phenomena
where the distribution of photon numbers is governed by the fractional
Poisson distribution Eq.(\ref{eq17}).

An attempt to create a family of coherent states with involvement of the
Mittag-Leffler function can be found in \cite{Solomon}, where formal
substitution in Eq.(\ref{eq10}) instead of $n!$ its generalization in terms
of $\Gamma (\alpha n+\beta )$, $(\alpha ,\beta >0)$ has been implemented. To
provide normalization condition the factor $e^{-\frac{1}{2}|z|^{2}}$ in Eq.(%
\ref{eq10}) has to be updated with ($E_{\alpha ,\beta }(|z|^{2}))^{-1/2}$,
where an entire function $E_{\alpha ,\beta }(|z|^{2})$ is generalization of
the Mittag-Leffler function (see, for details \cite{Solomon}).

Let's answer the question if the newly introduced coherent states $%
|\varsigma >$ are really generalized coherent states?

Quantum mechanical states are generalized coherent states if they \cite%
{Klauder1}:

(i) are parameterized continuously and normalized;

(ii) admit a resolution of unity with a positive weight function;

(iii) provide temporal stability, that is, the time evolving coherent state
belongs to the family of coherent states.

Let's now show that the new coherent states $|\varsigma >$ introduced by Eq.(%
\ref{eq14}) satisfy all above listed conditions.

To prove (i), we note that the coherent states $|\varsigma >$ are evidently
parametrized continuously by their label $\varsigma $ which is a complex
number $\varsigma =\xi +i\eta $, with $\xi =\mathrm{Re}\varsigma $ and $\eta
=\mathrm{Im}\varsigma $. Because of the normalization condition of the
fractional Poisson probability distribution $\sum\limits_{n=0}^{\infty
}P_{\mu }(n)=1$ and $<n|n^{\prime }>=\delta _{n,n^{\prime }}$, we have

\begin{equation*}
<\varsigma |\varsigma >=
\end{equation*}

\begin{equation}
\sum\limits_{n=0}^{\infty }\sum\limits_{n^{\prime }=0}^{\infty }<n|\frac{(%
\sqrt{\mu }\varsigma ^{\ast \mu })^{n}}{\sqrt{n!}}(E_{\mu }^{(n)}(-\mu
|\varsigma |^{2\mu }))^{1/2}\frac{(\sqrt{\mu }\varsigma ^{\mu })^{n^{\prime
}}}{\sqrt{n^{\prime }!}}(E_{\mu }^{(n^{^{\prime }})}(-\mu |\varsigma |^{2\mu
}))^{1/2}|n^{\prime }>=  \label{eq18}
\end{equation}

\begin{equation*}
\sum\limits_{n=0}^{\infty }\frac{(\mu |\varsigma |^{2\mu })^{n}}{n!}\left(
E_{\mu }^{(n)}(-\mu |\varsigma |^{2\mu })\right) =\sum\limits_{n=0}^{\infty
}P_{\mu }(n)=1,
\end{equation*}

that is, the coherent states $|\varsigma >$ are normalized.

To prove (ii), that is, the coherent states $|\varsigma >$ admit a
resolution of unity with a positive weight function, we introduce a positive
function $W_{\mu }(|\varsigma |^{2})>0$ which obeys the equation

\begin{equation}
\frac 1\pi \int\limits_Cd^2\varsigma |\varsigma >W_\mu (|\varsigma
|^2)<\varsigma |=I,  \label{eq19}
\end{equation}

where $d^2\varsigma =d(\mathrm{Re}\varsigma )d(\mathrm{Im}\varsigma )$ and
the integration extends over the entire complex plane $C$. This equation
with yet unknown function $W_\mu (|\varsigma |^2)$ can be considered as a
resolution of unity. To find the function $W_\mu (|\varsigma |^2)$ let's
transform Eq.(\ref{eq19}). Introducing new integration variables $\rho $ and 
$\phi $ by $\varsigma =\rho e^{i\phi }$, $d^2\varsigma =\rho ^2d\rho d\phi $
and making use of Eqs.(\ref{eq14}) and (\ref{eq15}) yield

\begin{equation*}
\frac 1\pi \int\limits_Cd^2\varsigma |\varsigma >W_\mu (|\varsigma
|^2)<\varsigma |=
\end{equation*}

\begin{equation}
\frac{1}{\pi }\sum\limits_{n=0}^{\infty }\sum\limits_{m=0}^{\infty
}\int\limits_{0}^{\infty }d\rho \rho ^{(n+m)\mu +1}\int\limits_{0}^{2\pi
}d\phi e^{i(n-m)\phi }\frac{W_{\mu }(\rho ^{2})}{\sqrt{n!m!}}\left( E_{\mu
}^{(n)}(-\mu |\rho |^{2\mu })\right) ^{1/2}\times  \label{eq20}
\end{equation}

\begin{equation*}
\left( E_\mu ^{(m)}(-\mu |\rho |^{2\mu })\right) ^{1/2}|n><m|=I.
\end{equation*}

By interchanging the orders of summation and integration and carrying out
the integration over $\phi $, we get a factor $2\pi \delta _{n,m}$, which
reduces the double summation to a single one. Therefore, Eq.(\ref{eq20}) is
simplified to

\begin{equation*}
\frac 1\pi \int\limits_Cd^2\varsigma |\varsigma >W_\mu (|\varsigma
|^2)<\varsigma |=
\end{equation*}

\begin{equation}
\sum\limits_{n=0}^{\infty }\frac{2}{n!}\int\limits_{0}^{\infty }d\rho \rho
^{2n\mu +1}W_{\mu }(\rho ^{2})E_{\mu }^{(n)}(-\mu |\rho |^{2\mu })|n><n|=I.
\label{eq21}
\end{equation}

Because of the completeness of orthonormal vectors $|n>$

\begin{equation}
\sum\limits_{n=0}^{\infty }|n><n|=I,  \label{eq22}
\end{equation}

we come to the following integral equation to find the positive function $%
W_{\mu }(x)$

\begin{equation}
\int\limits_0^\infty dxx^{\mu n}W_\mu (x)E_\mu ^{(n)}(-\mu x^\mu )=n!.
\label{eq23}
\end{equation}

To solve Eq.(\ref{eq23}) we use the Laplace transform of the function $%
t^{\mu n}E_{\mu }^{(n)}(-\mu t^{\mu })$, see Appendix,

\begin{equation}
\int\limits_{0}^{\infty }dte^{-st}t^{\mu n}E_{\mu }^{(n)}(-\mu t^{\mu })=%
\frac{n!\cdot s^{\mu -1}}{(s^{\mu }+\mu )^{n+1}}.  \label{eq24}
\end{equation}

By comparing Eqs.(\ref{eq23}) and (\ref{eq24}) we conclude that the positive
function $W_{\mu }(x)$ has the form

\begin{equation}
W_{\mu }(x)=(1-\mu )^{\frac{1-\mu }{\mu }}\cdot \exp \{-(1-\mu )^{1/\mu
}x\},\qquad 0<\mu \leq 1.  \label{eq25}
\end{equation}

Thus, we proved that the coherent states $|\varsigma >$ admit a resolution
of unity with the positive weight function $W_{\mu }(x)$ given by Eq.(\ref%
{eq25}). At $\mu =1$, function $W_{\mu }(x)$ becomes

\begin{equation*}
W_{1}(x)=\underset{\mu \rightarrow 1}{\lim }W_{\mu }(x)=\underset{\mu
\rightarrow 1}{\lim }\left[ (1-\mu )^{\frac{1-\mu }{\mu }}\cdot \exp
\{-(1-\mu )^{1/\mu }x\}\right] =1,
\end{equation*}

and we come back to the resolution of unity equation for the standard
coherent states $|z>$

\begin{equation}
\frac 1\pi \int\limits_Cd^2z|z><z|=I.  \label{eq26}
\end{equation}

To prove (iii), we note that if $|n>$ is an eigenvector of the Hamiltonian $%
H=\hbar \omega N=\hbar \omega a^{+}a$, where $\hbar $ is Planck's constant,
then the time evolution operator $\exp (-iHt/\hbar )$ results

\begin{equation*}
\exp (-iHt/\hbar )|n>=e^{-i\omega nt}|n>.
\end{equation*}

In other words, the time evolution of $|n>$ results in appearance of the
phase factor only while the state does not change. Let's consider time
evolution of the coherent state $|\varsigma >$ defined by Eq.(\ref{eq14}).
Well, as far as the coherent state is not an eigenstate of $H$ then one may
expect that it evolves into other states in time. However, it follows that

\begin{equation}
\exp (-iHt/\hbar )|\varsigma >=\sum\limits_{n=0}^{\infty }\frac{(\sqrt{\mu }%
\varsigma ^{\mu })^{n}}{\sqrt{n!}}(E_{\mu }^{(n)}(-\mu |\varsigma |^{2\mu
}))^{1/2}e^{-i\omega nt}|n>=|e^{-\frac{i\omega t}{\mu }}\varsigma >,
\label{eq27}
\end{equation}

which is just another coherent state belonging to a complex number $%
\varsigma e^{-\frac{i\omega t}\mu }$. We see that the time evolution of the
coherent state $|\varsigma >$ remains within the family of coherent states $%
|\varsigma >$. The property embodied in Eq.(\ref{eq27}) is the temporal
stability of coherent states $|\varsigma >$ under the action of $H$.

Thus, we conclude that the new coherent states $|\varsigma >$ satisfy the
Klauder's criteria set (i) - (iii)\ for generalized coherent states.

Finally, let us introduce an alternative notation for $|\varsigma >$ in
terms of the real $\xi $ and imaginary $\eta $ parts of $|\varsigma >$, that
is, $|\varsigma >=|\xi +i\eta >/\sqrt{2\hbar }$. Then from Eq.(\ref{eq14})
we have

\begin{equation}
|\varsigma >=|\xi ,\eta >=\sum\limits_{n=0}^{\infty }\frac{(\sqrt{\mu }(\xi
+i\eta )^{\mu })^{n}}{\sqrt{(2\hbar )^{n\mu }n!}}(E_{\mu }^{(n)}(-\mu \left( 
\frac{\xi ^{2}+\eta ^{2}}{2\hbar }\right) ^{\mu }))^{1/2}|n>.  \label{eq28}
\end{equation}

The adjoint coherent states are defined as

\begin{equation}
<\varsigma |=<\xi ,\eta |=\sum\limits_{n=0}^{\infty }<n|\frac{(\sqrt{\mu }%
(\xi -i\eta )^{\mu })^{n}}{\sqrt{(2\hbar )^{n\mu }n!}}(E_{\mu }^{(n)}(-\mu
\left( \frac{\xi ^{2}+\eta ^{2}}{2\hbar }\right) ^{\mu }))^{1/2}.
\label{eq29}
\end{equation}

Despite the fact that the adjoint state is labelled by $\varsigma $, the
series expansion Eq.(\ref{eq29}) are formed in fact of powers of $\varsigma
^{\ast }$.

\subsection{Quantum mechanical vector and operator representations based on
coherent states $|\protect\varsigma >$}

In spirit of Klauder's consideration \cite{Klauder}, let's show that the
resolution of unity criteria Eq.(\ref{eq19}) with $W_{\mu }(x)$ given by Eq.(%
\ref{eq25}) allows us to list fundamental quantum mechanical statements
pertaining to the associated representation of Hilbert space. Indeed, it is
easy to see that the newly introduced coherent states $|\varsigma >$ provide:

1. Inner Product of quantum mechanical vectors\textit{\ }$|\varphi >$\textit{%
\ }and\textit{\ }$|\psi >$ defined as

\begin{equation}
<\varphi |\psi >=\frac{1}{\pi }\int\limits_{C}d^{2}\varsigma <\varphi
|\varsigma >W_{\mu }(|\varsigma |^{2})<\varsigma |\psi >,  \label{eq30}
\end{equation}

where $d^{2}\varsigma =d(\mathrm{Re}\varsigma )d(\mathrm{Im}\varsigma )$ and
the integration extends over the entire complex plane $C$, weight function $%
W_{\mu }(x)$ is defined by Eq.(\ref{eq25}), the vector representatives are
wave functions $<\varphi |\varsigma >$ and $<\varsigma |\psi >$ given by 
\begin{equation}
<\varphi |\varsigma >=\sum\limits_{n=0}^{\infty }\frac{(\sqrt{\mu }\varsigma
^{\mu })^{n}}{\sqrt{n!}}(E_{\mu }^{(n)}(-\mu |\varsigma |^{2\mu
}))^{1/2}<\varphi |n>,  \label{eq31}
\end{equation}

\begin{equation}
<\varsigma |\psi >=\sum\limits_{n=0}^{\infty }<n|\psi >\frac{(\sqrt{\mu }%
\varsigma ^{\ast \mu })^{n}}{\sqrt{n!}}(E_{\mu }^{(n)}(-\mu |\varsigma
|^{2\mu }))^{1/2}.  \label{eq32}
\end{equation}

2. Vectors Transformation Law

\begin{equation}
<\varsigma |\mathcal{A}|\psi >=\frac 1\pi \int\limits_Cd^2\varsigma
^{^{\prime }}<\varsigma |\mathcal{A}|\varsigma ^{^{\prime }}>W_\mu
(|\varsigma ^{^{\prime }}|^2)<\varsigma ^{^{\prime }}|\psi >,  \label{eq33}
\end{equation}

where $<\varsigma |\mathcal{A}|\varsigma ^{^{\prime }}>$ is the matrix
element of quantum mechanical operator $\mathcal{A}$.

3. Operator Transformation Law

\begin{equation}
<\varsigma |\mathcal{A}_1\mathcal{A}_2|\varsigma ^{^{\prime }}>=\frac 1\pi
\int\limits_Cd^2\varsigma ^{^{^{\prime \prime }}}<\varsigma |\mathcal{A}%
_1|\varsigma ^{^{^{\prime \prime }}}>W_\mu (|\varsigma ^{^{\prime \prime
}}|^2)<\varsigma ^{^{\prime \prime }}|\mathcal{A}_2|\varsigma ^{^{\prime }}>,
\label{eq34}
\end{equation}

where $\mathcal{A}_{1}$ and $\mathcal{A}_{2}$ are two quantum mechanical
operators.

Further, the inverse map from the functional Hilbert space representation of
coherent states $|\varsigma >$ to the abstract one is provided by the
following decomposition laws:

4. Vector Decomposition Law

\begin{equation}
|\psi >=\frac{1}{\pi }\int\limits_{C}d^{2}\varsigma \ |\varsigma >W_{\mu
}(|\varsigma |^{2})<\varsigma |\psi >.  \label{eq35}
\end{equation}

5. Operator Decomposition Law

\begin{equation}
\mathcal{A}=\frac{1}{\pi }\int\limits_{C}d^{2}\varsigma _{1}d^{2}\varsigma
_{2}\ |\varsigma _{1}>W_{\mu }(|\varsigma _{1}|^{2})<\varsigma _{1}|\mathcal{%
A}|\varsigma _{2}>W_{\mu }(|\varsigma _{2}|^{2})<\varsigma _{2}|.
\label{eq36}
\end{equation}

Thus, we conclude that the resolution of unity Eq.(\ref{eq19}) with $W_{\mu
}(x)$, given by Eq.(\ref{eq25}), provides an appropriate inner product Eq.(%
\ref{eq30}) and lets us introduce the Hilbert space, Eqs.(\ref{eq33}) - (\ref%
{eq36}).

All of the above listed results lead to the well-know fundamental equations
of quantum optics and coherent states theory \cite{Klauder}, \cite{Glauber}
in the limit case $\mu =1$.

Table 2 summarizes the definitions and equations attributed to the newly
introduced coherent states $|\varsigma >$ with those for the coherent states 
$|z>$. Table 2 presents two sets of equations for a coherent state $|...>$,
for an adjoint coherent state \TEXTsymbol{<}$...|$, for the probability $%
P(n) $ that $n$ photons will be found in the coherent state $|...>$, for a
positive weight function $W(x)$ in the resolution of unity equations (\ref%
{eq19}) and (\ref{eq26}), the mean number $<...|a^{+}a|...>$ of photons when
the state is a coherent state $|...>$, and the quantum mechanical vector
decomposition law.

\begin{tabular}{|c|c|c|}
\hline
& $|\varsigma >$ $(0<\mu \leq 1)$ & $|z>$ \\ \hline
$|...>$ & $\sum\limits_{n=0}^{\infty }\frac{(\sqrt{\mu }\varsigma ^{\mu
})^{n}}{\sqrt{n!}}(E_{\mu }^{(n)}(-\mu |\varsigma |^{2\mu }))^{1/2}|n>$ & $%
e^{-\frac{1}{2}|z|^{2}}\sum\limits_{n=0}^{\infty }\frac{z^{n}}{\sqrt{n!}}|n>$
\\ \hline
\TEXTsymbol{<}$...|$ & $\sum\limits_{n=0}^{\infty }<n|\frac{(\sqrt{\mu }%
\varsigma ^{\ast \mu })^{n}}{\sqrt{n!}}(E_{\mu }^{(n)}(-\mu |\varsigma
|^{2\mu }))^{1/2}$ & \TEXTsymbol{<}$n|e^{-\frac{1}{2}|z|^{2}}\sum%
\limits_{n=0}^{\infty }\frac{z^{\ast n}}{\sqrt{n!}}$ \\ \hline
$P(n)$ & $\frac{(\mu |\varsigma |^{2\mu })^{n}}{n!}E_{\mu }^{(n)}(-\mu
|\varsigma |^{2\mu })$ & $\frac{|z|^{2n}}{n!}e^{-|z|^{2}}$ \\ \hline
$W(x)$ & $(1-\mu )^{\frac{1-\mu }{\mu }}\cdot \exp \{-(1-\mu )^{\frac{1}{\mu 
}}x\}$ & $1$ \\ \hline
$<...|a^{+}a|...>$ & $(\mu |\varsigma |^{2\mu })/\Gamma (\mu +1)$ & $|z|^{2}$
\\ \hline
$|\psi >$ & $\frac{1}{\pi }\int\limits_{C}d^{2}\varsigma |\varsigma >W_{\mu
}(|\varsigma |^{2})<\varsigma |\psi >$ & $\frac{1}{\pi }\int%
\limits_{C}d^{2}z|z><z|\psi >$ \\ \hline
\end{tabular}

Table 2. \textit{Coherent states }$|\varsigma >$\textit{\ vs coherent states 
}$|z>$.

\section{Generalized Bell and Stirling Numbers}

\subsection{Fractional Bell polynomials and fractional Bell numbers}

Based on the fractional Poisson probability distribution Eq.(\ref{eq7}) we
introduce a new generalization of the Bell polynomials

\begin{equation}
B_{\mu }(x,m)=\sum\limits_{n=0}^{\infty }n^{m}\frac{x^{n}}{n!}%
\sum\limits_{k=0}^{\infty }\frac{(k+n)!}{k!}\frac{(-x)^{k}}{\Gamma (\mu
(k+n)+1)},\qquad B_{\mu }(x,0)=1,  \label{eq37}
\end{equation}

where the parameter $\mu $ is $0<\mu \leq 1.$ We will call $B_{\mu }(x,m)$
as the fractional Bell polynomials of $m$-order. A few fractional Bell
polynomials are

\begin{equation}
B_{\mu }(x,1)=\frac{x}{\Gamma (\mu +1)},  \label{eq37.1}
\end{equation}

\begin{equation}
B_{\mu }(x,2)=\frac{2x^{2}}{\Gamma (2\mu +1)}+\frac{x}{\Gamma (\mu +1)},
\label{eq37.2}
\end{equation}

\begin{equation}
B_{\mu }(x,3)=\frac{6x^{3}}{\Gamma (3\mu +1)}+\frac{6x^{2}}{\Gamma (2\mu +1)}%
+\frac{x}{\Gamma (\mu +1)},  \label{eq37.3}
\end{equation}

\begin{equation}
B_{\mu }(x,4)=\frac{24x^{4}}{\Gamma (4\mu +1)}+\frac{36x^{3}}{\Gamma (3\mu
+1)}+\frac{14x^{2}}{\Gamma (2\mu +1)}+\frac{x}{\Gamma (\mu +1)}.
\label{eq37.4}
\end{equation}

The polynomials $B_{\mu }(x,m)$ are related to the well-known Bell
polynomials \cite{Bell} $B(x,m)$ by

\begin{equation}
B_{\mu }(x,m)|_{\mu =1}=B(x,m)=e^{-x}\sum\limits_{n=0}^{\infty }n^{m}\frac{%
x^{n}}{n!}.  \label{eq38}
\end{equation}

From Eq.(\ref{eq37}) at $x=1$ we come to a new numbers $B_{\mu }(m)$, which
we call the fractional Bell numbers

\begin{equation}
B_{\mu }(m)=B_{\mu }(x,m)|_{x=1}=\sum\limits_{n=0}^{\infty }\frac{n^{m}}{n!}%
\sum\limits_{k=0}^{\infty }\frac{(k+n)!}{k!}\frac{(-1)^{k}}{\Gamma (\mu
(k+n)+1)}.  \label{eq39}
\end{equation}

As an example, here are a few fractional Bell numbers

\begin{equation*}
B_\mu (0)=1,\qquad B_\mu (1)=\frac 1{\Gamma (\mu +1)},\qquad B_\mu (2)=\frac
2{\Gamma (2\mu +1)}+\frac 1{\Gamma (\mu +1)},
\end{equation*}

\begin{equation*}
B_{\mu }(3)=\frac{6}{\Gamma (3\mu +1)}+\frac{6}{\Gamma (2\mu +1)}+\frac{1}{%
\Gamma (\mu +1)},
\end{equation*}

\begin{equation*}
B_{\mu }(4)=\frac{24}{\Gamma (4\mu +1)}+\frac{36}{\Gamma (3\mu +1)}+\frac{14%
}{\Gamma (2\mu +1)}+\frac{1}{\Gamma (\mu +1)}.
\end{equation*}

It is easy to see that Eq.(\ref{eq39}) can be written as

\begin{equation}
B_\mu (m)=\sum\limits_{n=0}^\infty \frac{n^m}{n!}E_\mu ^{(n)}(-1),
\label{eq40}
\end{equation}

where $E_{\mu }^{(n)}(-1)=(d^{n}E_{\mu }(z)/dz^{n})|_{z=-1}$ and $E_{\mu
}(z) $ is given by Eq.(\ref{eq6}). The new formula Eq.(\ref{eq40}) is in
fact a fractional generalization of so-called Dobi\'{n}ski relation known
since 1877 \cite{Dobinski}, \cite{Rota}. Indeed, at $\mu =1$ when the
Mittag-Leffler function is just the exponential function, $E_{1}(z)=\exp (z)$%
, we have, $E_{1}^{(n)}(-1)=(d^{n}E_{1}(z)/dz^{n})|_{z=-1}=e^{-1}$, and the
equation (\ref{eq40}) becomes the Dobi\'{n}ski relation \cite{Dobinski} for
the Bell numbers $B(m)$,

\begin{equation}
B(m)=B_{\mu }(m)|_{\mu =1}=e^{-1}\sum\limits_{n=0}^{\infty }\frac{n^{m}}{n!}.
\label{eq41}
\end{equation}

Now we focus on the general definitions given by Eqs.(\ref{eq37}) and (\ref%
{eq40}) to find the generating functions of the polynomials $B_{\mu }(x,m)$
and the numbers $B_{\mu }(m).$ Let us introduce the generating function of
the polynomials $B_{\mu }(x,m)$ as 
\begin{equation}
F_{\mu }(s,x)=\sum\limits_{m=0}^{\infty }\frac{s^{m}}{m!}B_{\mu }(x,m).
\label{eq42}
\end{equation}

Therefore, to get the polynomial $B_{\mu }(x,m)$ we should differentiate $%
F_{\mu }(s,x)$ $m$ times with respect to $s$, and then let $s=0$. That is, 
\begin{equation}
B_{\mu }(x,m)=\frac{\partial ^{m}}{\partial s^{m}}F_{\mu }(s,x)|_{s=0}.
\label{eq43}
\end{equation}

To find an explicit equation for $F_{\mu }(s,x)$, let's substitute Eq.(\ref%
{eq37}) into Eq.(\ref{eq42}) and evaluate the sum over $m$. As a result we
have

\begin{equation}
F_{\mu }(s,x)=\sum\limits_{n=0}^{\infty }\frac{x^{n}}{n!}e^{sn}\sum%
\limits_{k=0}^{\infty }\frac{(k+n)!}{k!}\frac{(-x)^{k}}{\Gamma (\mu (k+n)+1)}%
.  \label{eq44}
\end{equation}

Then, introducing the new summation variable $l=k+n$, yields 
\begin{equation*}
F_{\mu }(s,x)=\sum\limits_{n=0}^{\infty }\frac{x^{n}}{n!}e^{sn}\sum%
\limits_{l=n}^{\infty }\frac{l!}{(l-n)!}\frac{(-x)^{l-n}}{\Gamma (\mu l+1)}=
\end{equation*}

\begin{equation*}
\sum\limits_{l=0}^\infty \frac 1{\Gamma (\mu l+1)}\sum\limits_{n=0}^l\frac{l!%
}{n!(l-n)!}e^{sn}x^n(-x)^{l-n}=\sum\limits_{l=0}^\infty \frac{(xe^s-x)^l}{%
\Gamma (\mu l+1)}.
\end{equation*}

Finally, we obtain

\begin{equation}
F_{\mu }(s,x)=E_{\mu }(x(e^{s}-1)),  \label{eq45}
\end{equation}

where $E_{\mu }(z)$ is the Mittag-Leffler function defined by the power
series Eq.(\ref{eq6}).

Thus, we have

\begin{equation}
E_{\mu }(x(e^{s}-1))=\sum\limits_{m=0}^{\infty }\frac{s^{m}}{m!}B_{\mu
}(x,m).  \label{eq45.1}
\end{equation}

It is easy to see that the generating function $F_{\mu }(s,x)$, given by Eq.(%
\ref{eq45}), can be considered as the moment generating function of the
fractional Poisson probability distribution (see Eq.(35) in Ref.\cite%
{Laskin7}).

In the case of $\mu =1$, Eq.(\ref{eq45}) turns into the equation for the
generating function of the Bell polynomials $B(x,m)$ defined by Eq.(\ref%
{eq38}),

\begin{equation}
F_{1}(s,x)=\exp \{x(e^{s}-1)\}=\sum\limits_{m=0}^{\infty }\frac{s^{m}}{m!}%
B(x,m).  \label{eq46}
\end{equation}

If we put $x=1$ in Eq.(\ref{eq45}), then we immediately come to the
generating function $\mathcal{B}_{\mu }(s)$ of the fractional Bell numbers $%
B_{\mu }(m)$

\begin{equation}
\mathcal{B}_{\mu }(s)=\sum\limits_{m=0}^{\infty }\frac{s^{m}}{m!}B_{\mu
}(m)=E_{\mu }(e^{s}-1).  \label{eq47}
\end{equation}

The numbers $B_{\mu }(m)$ can be obtained by differentiating $\mathcal{B}%
_{\mu }(s)$ $m$ times with respect to $s$, and then letting $s=0$,

\begin{equation}
B_{\mu }(m)=\frac{\partial ^{m}}{\partial s^{m}}\mathcal{B}_{\mu
}(s,x)|_{s=0}.  \label{eq48}
\end{equation}

When $\mu =1$, Eq.(\ref{eq47}) reads

\begin{equation}
\mathcal{B}_1(s)=\sum\limits_{m=0}^\infty \frac{s^m}{m!}B_1(m)=\exp (e^s-1),
\label{eq49}
\end{equation}

and we come to the well-known equation for the generating function of the
Bell numbers.

Now we are going to consider quantum physics and probability theory problems
where the fractional Bell polynomials appear.

\subsubsection{Quantum physics application of the fractional Bell polynomials%
}

As quantum physics applications of the fractional Bell polynomials let us
show how the fractional Bell polynomials are related to the new coherent
states \TEXTsymbol{\vert}$\varsigma >$ introduced by Eq.(\ref{eq14}). For
the boson creation $a^{+}$ and $a$ annihilation operators of a photon field
that satisfy the commutation relation $[a,a^{+}]=aa^{+}-a^{+}a=\QTR{sl}{1}$,
the diagonal matrix element of the $n$-th power of the number operator $%
(a^{+}a)^{n}$ yields the fractional Bell polynomials of order $n$,

\begin{equation}
<\varsigma |(a^{+}a)^n|\varsigma >=B_\mu (|\varsigma |^{2\mu },n).
\label{eq50}
\end{equation}

Then, the diagonal coherent state $|\varsigma >$ matrix element $<\varsigma
|\exp (-iHt/\hbar )|\varsigma >$ of the time evolution operator

\begin{equation}
\exp (-iHt/\hbar )=\exp \left\{ (-i\omega t/\hbar )a^{+}a\right\}
\label{eq51}
\end{equation}

can be written as

\begin{equation}
<\varsigma |\exp \left\{ (-i\omega t/\hbar )a^{+}a\right\} |\varsigma
>=\sum\limits_{n=0}^{\infty }\frac{1}{n!}(-i\frac{\omega t}{\hbar }%
)^{n}<\varsigma |(a^{+}a)^{n}|\varsigma >=  \label{eq52}
\end{equation}

\begin{equation*}
\sum\limits_{n=0}^\infty \frac 1{n!}(-i\frac{\omega t}\hbar )^nB_\mu
(|\varsigma |^{2\mu },n),
\end{equation*}

where $\hbar $ is Planck's constant. By comparing with Eq.(\ref{eq42}) we
conclude that

\begin{equation}
<\varsigma |\exp \left\{ (-i\omega t/\hbar )a^{+}a\right\} |\varsigma
>=E_{\mu }\left( |\varsigma |^{2\mu }(\exp (-i\omega t/\hbar )-1)\right) .
\label{eq53}
\end{equation}

In other words, the diagonal coherent state $|\varsigma >$ matrix element of
the operator $\exp \left\{ (-i\omega t/\hbar )a^{+}a\right\} $ is the
generating function of the fractional Bell polynomials. In the special case $%
\mu =1$, Eq.(\ref{eq53}) reads

\begin{equation}
<z|\exp \left\{ (-i\omega t/\hbar )a^{+}a\right\} |z>=\exp \{|z|^{2}(\exp
(-i\omega t/\hbar )-1)\},  \label{eq54}
\end{equation}

that is, the diagonal coherent state $|z>$ matrix element of the operator $%
\exp \left\{ (-i\omega t/\hbar )a^{+}a\right\} $ is the generating function
of the Bell polynomials. This statement immediately follows from
Eqs.(11.5-2) and (11.2-10) of Ref.\cite{Wolf} for the diagonal matrix
element of the operator $\exp \left\{ (-i\omega t/\hbar )a^{+}a\right\} $ in
the basis of the coherent states $|z>$.

It follows from Eq.(\ref{eq50}) that for the special case when $|\varsigma
|=1$ the diagonal matrix element of the $n$-th power of the number operator $%
(a^{+}a)^{n}$ yields the fractional Bell number of order $n$,

\begin{equation}
<\varsigma |(a^{+}a)^{n}|\varsigma >|_{|\varsigma |=1}=B_{\mu }(|\varsigma
|^{2\mu },n)|_{|\varsigma |=1}=B_{\mu }(n).  \label{eq50a}
\end{equation}

At $\mu =1$ this equation turns into the relationship between the well-known
Bell numbers $B(n)$ and the diagonal matrix element of the $n$-th power of
the number operator $(a^{+}a)^{n}$ in the basis of the standard coherent
states $|z>$,

\begin{equation}
<z|(a^{+}a)^{n}|z>|_{|z|=1}=B(|z|^{2\mu },n)|_{|z|=1}=B(n).  \label{eq50b}
\end{equation}

The equation (\ref{eq50b}) originally was obtained in \cite{Katriel}.

\subsubsection{Fractional compound Poisson processes}

As another example where the fractional Bell polynomials come from, we
consider the\textit{\ fractional compound Poisson process} first introduced
into the probability theory and developed by Laskin in \cite{Laskin7}. Let
us consider the pair $\{N(t)$, $Y_{i}\}$, where $\{N(t)$, $t\geq 0]\}$ is a
counting Poisson process with a probability distribution function $P(n,t)$
and $\{Y_{i}$, $i=1,2,...\}$ is a family of independent and identically
distributed discrete random variables with probability distribution function 
$p(Y)$ for each $Y_{i}$. The process $\{N(t)$, $t\geq 0\}$ and the sequence $%
\{Y_{i}$, $i=1,2,...\}$ are assumed to be independent. To be more specific
we will distinguish the following four cases:

1. In the pair $\{N(t)$, $Y_{i}\}$ the counting process $N(t)$ is the
fractional Poisson process with probability distribution function given by
Eq.(\ref{eq7}) and $Y_{i}$ are random variables with the fractional Poisson
probability distribution, that is the probability that $Y=n$ has a form

\begin{equation}
p(Y=n)=\frac{(\lambda _{\mu })^{n}}{n!}\sum\limits_{k=0}^{\infty }\frac{%
(k+n)!}{k!}\frac{(-\lambda _{\mu })^{k}}{\Gamma (\mu (k+n)+1)},\qquad 0<\mu
\leq 1,  \label{eq50ff}
\end{equation}

where $\lambda _{\mu }$ is the parameter associated with random variable $Y$.

2. In the pair $\{N(t)$, $Y_{i}\}$ the counting process $N(t)$ is the
fractional Poisson process with probability distribution function given by
Eq.(\ref{eq7}) and $Y_{i}$ are random variables with the Poisson probability
distribution, that is the probability that $Y=n$ has a form

\begin{equation}
p(Y=n)=\frac{(\lambda )^{n}}{n!}e^{-\lambda }.  \label{eq50fp}
\end{equation}

3. In the pair $\{N(t)$, $Y_{i}\}$ the counting process $N(t)$ is the
Poisson process with probability distribution function given by Eq.(\ref%
{eq9.1}) and $Y_{i}$ are random variables with the fractional Poisson
probability distribution, that is the probability that random variable $Y=n$
is given by Eq.(\ref{eq50ff}).

4. In the pair $\{N(t)$, $Y_{i}\}$ the counting process $N(t)$ is the
Poisson process with probability distribution function given by Eq.(\ref%
{eq9.1}) and $Y_{i}$ are random variables with the Poisson probability
distribution, that is the probability that $Y=n$ is given by Eq.(\ref{eq50fp}%
).

The compound Poisson process $X(t)$ is represented by

\begin{equation}
X(t)=\sum\limits_{i=1}^{N(t)}Y_{i}.  \label{eq51f}
\end{equation}

The moment generating function $J_{\mu }(s,t)$ of compound Poisson process
has been introduced as \cite{Laskin7}

\begin{equation}
J_{\mu }(s,t)=<\exp \{sX(t)\}>_{N(t),Y_{i}},  \label{eq52f}
\end{equation}

where $<...>_{N(t),Y_{i}}$ stands for two statistically independent
averaging procedures:

a). Averaging over random number $n$ governed by the counting Poisson process

\begin{equation}
<...>_{N(t)}=\sum\limits_{n=0}^{\infty }P(n,t)...,  \label{eq53f}
\end{equation}

where $P(n,t)$ is given either by Eq.(\ref{eq7}) or by Eq.(\ref{eq9.1}).

b). Averaging over independent random variables $Y_{i}$, $<...>_{Y_{i}}$

\begin{equation}
<...>_{Y_{i}}=\int dY_{1}...dY_{n}p(Y_{1})...p(Y_{n})...,  \label{eq54f}
\end{equation}

where $p(Y_{i})$ is the probability density of random variable $Y_{i}$ given
either by Eq.(\ref{eq50ff}) or by Eq.(\ref{eq50fp}).

To obtain equation for the moment generating function $J_{\mu }(s,t)$ we
apply Eqs.(\ref{eq53f}) and (\ref{eq54f}) to Eq.(\ref{eq52f}),

\begin{equation*}
J_{\mu }(s,t)=\sum\limits_{n=0}^{\infty }P_{\mu }(n,t)<\exp
\{sX(t)|N(t)=n\}>_{Y_{i}}=
\end{equation*}

\begin{equation}
\sum\limits_{n=0}^{\infty }<\exp \{s(Y_{1}+...+Y_{n})\}>_{Y_{i}}\times \frac{%
(\nu t^{\mu })^{n}}{n!}\sum\limits_{k=0}^{\infty }\frac{(k+n)!}{k!}\frac{%
(-\nu t^{\mu })^{k}}{\Gamma (\mu (k+n)+1)}=  \label{eq55fg}
\end{equation}

\begin{equation*}
\sum\limits_{n=0}^{\infty }<\exp \{s(Y_{1})\}>_{Y_{i}}^{n}\times \frac{(\nu
t^{\mu })^{n}}{n!}\sum\limits_{k=0}^{\infty }\frac{(k+n)!}{k!}\frac{(-\nu
t^{\mu })^{k}}{\Gamma (\mu (k+n)+1)},
\end{equation*}

where we used the independence between $N(t)$ and $\{Y_{1},Y_{2},...\}$, and
the independence of the $Y_{i}$'s between themselves. Hence, letting

\begin{equation}
g(s)=<e^{sY}>_{Y},  \label{eq56fg}
\end{equation}

for the moment generating function of random variables $Y_{i}$, we find from
Eq.(\ref{eq55fg}) the moment generating function $J_{\mu }(s,t)$ of the
fractional compound Poisson process

\begin{equation}
J_{\mu }(s,t)=\sum\limits_{n=0}^{\infty }g^{n}(s)\times \frac{(\nu t^{\mu
})^{n}}{n!}\sum\limits_{k=0}^{\infty }\frac{(k+n)!}{k!}\frac{(-\nu t^{\mu
})^{k}}{\Gamma (\mu (k+n)+1)}=E_{\mu }(\nu t^{\mu }(g(s)-1)).  \label{eq58f}
\end{equation}

It follows from Eq.(\ref{eq52f}) that $k^{\mathrm{th}}$ order moment of
fractional compound Poisson process $X(t)$ is obtained by differentiating $%
J_{\mu }(s,t)$ $k$ times with respect to $s$, then putting $s=0$, that is

\begin{equation}
<X^{k}(t)>_{N(t),Y_{i}}=\frac{\partial ^{k}}{\partial s^{k}}J_{\mu
}(s,t)_{|s=0}.  \label{eq55f1}
\end{equation}

For the case 1, when $N(t)$ is the fractional Poisson process and $Y_{i}$
are random variables with the fractional Poisson probability distribution,
we obtain

\begin{equation}
J_{\mu }^{(1)}(s,t)=E_{\mu }(\nu t^{\mu }(g_{1}(s)-1)),  \label{eq55ff}
\end{equation}

with

\begin{equation}
g_{1}(s)=<e^{sY}>_{Y_{i}}=E_{\mu }(\lambda _{\mu }(e^{s}-1)).
\label{eq551ff}
\end{equation}

Thus, we have

\begin{equation}
J_{\mu }^{(1)}(s,t)=E_{\mu }(\nu t^{\mu }\{E_{\mu }(\lambda _{\mu
}(e^{s}-1))-1\}).  \label{eq552ff}
\end{equation}

For the case 2, when $N(t)$ is the fractional Poisson process and $Y_{i}$
are random variables with the Poisson probability distribution, we obtain

\begin{equation}
J_{\mu }^{(2)}(s,t)=E_{\mu }(\nu t^{\mu }(g_{2}(s)-1)),  \label{eq55fp}
\end{equation}

with

\begin{equation}
g_{2}(s)=<e^{sY}>_{Y_{i}}=\exp (\lambda (e^{s}-1)).  \label{eq551fp}
\end{equation}

Thus, we have

\begin{equation}
J_{\mu }^{(2)}(s,t)=E_{\mu }(\nu t^{\mu }\{\exp (\lambda (e^{s}-1))-1\}).
\label{eq552fp}
\end{equation}

For the case 3, when $N(t)$ is the fractional Poisson process and $Y_{i}$
are random variables with the fractional Poisson probability distribution,
we obtain

\begin{equation}
J_{\mu }^{(3)}(s,t)=\exp (\overline{\nu }t(g_{3}(s)-1)),  \label{eq55pf}
\end{equation}

with

\begin{equation}
g_{3}(s)=<e^{sY}>_{Y_{i}}=E_{\mu }(\lambda _{\mu }(e^{s}-1)).
\label{eq551pf}
\end{equation}

and the parameter $\overline{\nu }$ comes from Eq.(\ref{eq9.1}).

Thus, we have

\begin{equation}
J_{\mu }^{(3)}(s,t)=\exp (\overline{\nu }t\{E_{\mu }(\lambda _{\mu
}(e^{s}-1))-1\}).  \label{eq552pf}
\end{equation}

For the case 4, when $N(t)$ is the Poisson process and $Y_{i}$ are random
variables with the Poisson probability distribution, we obtain

\begin{equation}
J^{(4)}(s,t)=\exp (\overline{\nu }t(g_{4}(s)-1)),  \label{eq55pp}
\end{equation}

with

\begin{equation}
g_{4}(s)=<e^{sY}>_{Y_{i}}=\exp (\lambda (e^{s}-1)).  \label{eq551pp}
\end{equation}

Thus, we have

\begin{equation}
J^{(4)}(s,t)=\exp (\overline{\nu }t\{\exp (\lambda (e^{s}-1))-1\}).
\label{eq552pp}
\end{equation}

Now, let us take a look at the moment generating function $J_{\mu
}^{(2)}(s,t)$ given by Eq.(\ref{eq552fp}). It can be expanded as

\begin{equation}
J_{\mu }^{(2)}(s,t)=\sum\limits_{m=0}^{\infty }\frac{(\lambda (e^{s}-1))^{m}%
}{m!}B_{\mu }(\nu t^{\mu },m)=  \label{eq5532}
\end{equation}

\begin{equation}
\sum\limits_{m=0}^{\infty }\lambda ^{m}\sum\limits_{k=m}^{\infty }\frac{s^{k}%
}{k!}S(k,m)B_{\mu }(\nu t^{\mu },m).
\end{equation}

where we used Eqs (\ref{eq45.1}) and (\ref{eq72}). Finally, we have for $%
J_{\mu }^{(2)}(s,t)$,

\begin{equation}
J_{\mu }^{(2)}(s,t)=E_{\mu }(\nu t^{\mu }\{\exp (\lambda
(e^{s}-1))-1\})=\sum\limits_{k=0}^{\infty }\frac{s^{k}}{k!}%
\sum\limits_{m=0}^{k}\lambda ^{m}S(k,m)B_{\mu }(\nu t^{\mu },m).
\label{eq5531}
\end{equation}

Then as it follows from Eq.(\ref{eq55f1}) that $k^{\mathrm{th}}$ order
moment of fractional compound Poisson process for the case 2 is

\begin{equation}
<X^{k}(t)>_{N(t),Y_{i}}^{(2)}=\frac{\partial ^{k}}{\partial s^{k}}J_{\mu
}^{(2)}(s,t)_{|s=0}=\sum\limits_{m=1}^{k}\lambda ^{m}S(k,m)B_{\mu }(\nu
t^{\mu },m),  \label{eq554}
\end{equation}

where $S(k,m)$ are the standard Stirling numbers of the second kind and $%
B_{\mu }(\nu t^{\mu },m)$ are the fractional Bell polynomials defined by Eq.(%
\ref{eq37}).

Thus, this example shows the way the fractional Bell polynomials appear, if
one evaluates moments of fractional compound Poisson process for the case 2.

It is easy to see that when $\mu =1$, the case 2 turns into the case 4.
Hence, at $\mu =1$ for the moment generating function $J^{(4)}(s,t)$ we have

\begin{equation}
J^{(4)}(s,t)=\sum\limits_{k=0}^{\infty }\frac{s^{k}}{k!}\sum%
\limits_{m=0}^{k}\lambda ^{m}S(k,m)B(\overline{\nu }t,m),  \label{eq5541}
\end{equation}

and $k^{\mathrm{th}}$ order moment of compound Poisson process for the case
4 is

\begin{equation}
<X^{k}(t)>_{N(t),Y_{i}}^{(4)}=\frac{\partial ^{k}}{\partial s^{k}}%
J^{(4)}(s,t)_{|s=0}=\sum\limits_{m=1}^{k}\lambda ^{m}S(k,m)B(\overline{\nu }%
t,m),  \label{5542}
\end{equation}

where $B(\nu t^{\mu },m)$ are the Bell polynomials defined by Eq.(\ref{eq38}%
).

The first two moments of fractional compound Poisson process for the case 2
are

\begin{equation}
<X^{1}(t)>_{N(t),Y_{i}}^{(2)}=\frac{\partial }{\partial s}J_{\mu
}^{(2)}(s,t)_{|s=0}=\lambda S(1,1)B_{\mu }(\nu t^{\mu },1)=\lambda \frac{\nu
t^{\mu }}{\Gamma (\mu +1)},  \label{eq5543}
\end{equation}

\begin{equation}
<X^{2}(t)>_{N(t),Y_{i}}^{(2)}=\frac{\partial ^{2}}{\partial s^{2}}J_{\mu
}^{(2)}(s,t)_{|s=0}=\sum\limits_{m=1}^{2}\lambda ^{m}S(2,m)B_{\mu }(\nu
t^{\mu },m),  \label{eq5544}
\end{equation}

Equation (\ref{eq5543}) for the first order moment of fractional Poisson
process was found at first time in \cite{Laskin7}. Using expressions for
fractional Bell polynomials $B_{\mu }(\nu t^{\mu },1)$ and $B_{\mu }(\nu
t^{\mu },2)$ yields

\begin{equation}
<X^{2}(t)>_{N(t),Y_{i}}^{(2)}=\lambda ^{2}\left( \frac{2(\nu t^{\mu })^{2}}{%
\Gamma (2\mu +1)}+\frac{\nu t^{\mu }}{\Gamma (\mu +1)}\right) +\lambda \frac{%
\nu t^{\mu }}{\Gamma (\mu +1)}.  \label{eq5544.1}
\end{equation}

with $S(2,1)=1$ and $S(2,2)=1$.

Then, the variance $\sigma _{N(t),Y_{i}}^{(2)}$ for the case 2 is

\begin{equation*}
\sigma _{N(t),Y_{i}}^{(2)}=\left(
<X^{2}(t)>_{N(t),Y_{i}}^{(2)}-(<X^{1}(t)>_{N(t),Y_{i}}^{(2)})^{2}\right) =
\end{equation*}

\begin{equation*}
\lambda ^{2}\left( \frac{2(\nu t^{\mu })^{2}}{\Gamma (2\mu +1)}+\frac{\nu
t^{\mu }}{\Gamma (\mu +1)}\right) +\lambda \frac{\nu t^{\mu }}{\Gamma (\mu
+1)}-\left( \lambda \frac{\nu t^{\mu }}{\Gamma (\mu +1)}\right) ^{2}.
\end{equation*}

Or, after simple transformations (see, for instance, page 207 in \cite%
{Laskin7}) we have

\begin{equation}
\sigma _{N(t),Y_{i}}^{(2)}=\lambda ^{2}\left( \frac{\mu B(\mu ,\frac{1}{2})}{%
2^{\mu -1}}-1\right) \left( \frac{\nu t^{\mu }}{\Gamma (\mu +1)}\right)
^{2}+(\lambda ^{2}+\lambda )\frac{\nu t^{\mu }}{\Gamma (\mu +1)},
\label{eq5544.4}
\end{equation}

where $B(\mu ,\frac{1}{2})$ is the Beta-function defined as\cite{Abramowitz3}

\begin{equation}
B(\mu ,\nu )=\frac{\Gamma (\mu )\Gamma (\nu )}{\Gamma (\mu +\nu )}.
\label{eq5544.5}
\end{equation}

At $\mu =1$ Eqs.(\ref{eq5543}) and (\ref{eq5544}) are transformed into the
first two moments of compound Poisson process for the case 4,

\begin{equation}
<X^{1}(t)>_{N(t),Y_{i}}^{(4)}=\frac{\partial }{\partial s}%
J^{(4)}(s,t)_{|s=0}=\lambda S(1,1)B(\overline{\nu }t,1)=\lambda \overline{%
\nu }t,  \label{eq5545}
\end{equation}

\begin{equation}
<X^{2}(t)>_{N(t),Y_{i}}^{(4)}=\frac{\partial ^{2}}{\partial s^{2}}%
J^{(4)}(s,t)_{|s=0}=\sum\limits_{m=1}^{2}\lambda ^{m}S(2,m)B(\overline{\nu }%
t,m)=  \label{eq5546}
\end{equation}

\begin{equation*}
\lambda ^{2}(\overline{\nu }t)^{2}+(\lambda ^{2}+\lambda )\overline{\nu }t.
\end{equation*}

Then the variance $\sigma _{N(t),Y_{i}}^{(4)}$ for the case 4 is

\begin{equation}
\sigma _{N(t),Y_{i}}^{(4)}=\left(
<X^{2}(t)>_{N(t),Y_{i}}^{(4)}-(<X^{1}(t)>_{N(t),Y_{i}}^{(4)})^{2}\right)
=(\lambda ^{2}+\lambda )\overline{\nu }t.  \label{eq5547}
\end{equation}

We see, that Eq.(\ref{eq5547}) follows straighforwardly from Eq.(\ref%
{eq5544.4}) at $\mu =1$ with substitution $\nu \rightarrow \overline{\nu }$.

The Table 3 displays the moment generating functions for each of four cases
introduced above.

\begin{tabular}{|l|l|l|}
\hline
& $N(t)$, $0<\mu <1$ & $N(t)$, $\mu =1$ \\ \hline
$Y_{i}$, $0<\mu <1$ & $\QATOP{{\Large Case}\text{ }{\Large 1}\QATOP{{}}{%
\QATOP{{}}{{}}}\qquad \qquad \qquad \qquad }{{\LARGE E}_{\mu }{\LARGE (\nu t}%
^{\mu }{\LARGE \{E}_{\mu }{\LARGE (\lambda }_{\mu }{\LARGE (e}^{s}{\LARGE %
-1))-1\})}\QATOP{{}}{\QATOP{{}}{{}}}}$ & $\QATOP{{\Large Case}\text{ }%
{\Large 3}\QATOP{{}}{\QATOP{{}}{{}}}\qquad \qquad \qquad \qquad }{\exp 
{\LARGE (}\overline{\nu }{\LARGE t\{E}_{\mu }{\LARGE (\lambda _{\mu }(e}^{s}%
{\LARGE -1))-1\})}\QATOP{{}}{\QATOP{{}}{{}}}}$ \\ \hline
$Y_{i}$, $\mu =1$ & $\QATOP{{\Large Case}\text{ }{\Large 2}\QATOP{{}}{\QATOP{%
{}}{{}}}\qquad \qquad \qquad \qquad }{{\LARGE E}_{\mu }{\LARGE (\nu t}^{\mu }%
{\LARGE \{}\exp {\LARGE (\lambda (e}^{s}{\LARGE -1))-1\})}\QATOP{{}}{\QATOP{%
{}}{{}}}}$ & $\QATOP{{\Large Case}\text{ }{\Large 4}\QATOP{{}}{\QATOP{{}}{{}}%
}\qquad \qquad \qquad \qquad }{\exp {\LARGE (}\overline{\nu }{\LARGE t\{}%
\exp {\LARGE (\lambda (e}^{s}{\LARGE -1))-1\})}\QATOP{{}}{\QATOP{{}}{{}}}}$
\\ \hline
\end{tabular}

Table 3. \textit{The moment generating functions of fractional compound
Poisson processes}.

\subsection{Fractional Stirling numbers of the second kind}

We introduce the fractional generalization of the Stirling numbers\footnote{%
Stirling numbers, introduced by J. Stirling \cite{Stirling} in 1730, have
been studied in the past by many celebrated mathematicians. Among them are
Euler, Lagrange, Laplace and Cauchy. Stirling numbers play an important role
in combinatorics, number theory, probability and statistics. There are two
common sets of Stirling numbers, they are so-called Stirling numbers of the
first kind and Stirling numbers of the second kind (for details, see Refs.%
\cite{Charalambides1}, \cite{Charalambides}).} of the second kind $S_{\mu
}(m,l)$ by means of equation

\begin{equation}
B_{\mu }(x,m)=\sum\limits_{l=0}^{m}S_{\mu }(m,l)x^{l},  \label{eq55}
\end{equation}

where $B_{\mu }(x,m)$ is a fractional generalization of the Bell polynomials
given by Eq.(\ref{eq37}) and the parameter $\mu $ is $0<\mu \leq 1$. At $\mu
=1$, Eq.(\ref{eq55}) defines the integers $S(m,l)=S_{\mu }(m,l)|_{\mu =1}$,
which are called Stirling numbers of the second kind. At $x=1$, when the
fractional Bell polynomials $B_{\mu }(x,m)$ become the fractional Bell
numbers, $B_{\mu }(m)=B_{\mu }(x,m)|_{x=1}$, Eq.(\ref{eq55}) gives us a new
equation to express fractional Bell numbers in terms of fractional Stirling
numbers of the second kind

\begin{equation}
B_{\mu }(m)=\sum\limits_{l=0}^{m}S_{\mu }(m,l).  \label{eq56}
\end{equation}

To find $S_\mu (m,l)$ we transform the right-hand side of Eq.(\ref{eq37}) as
follows

\begin{equation*}
B_{\mu }(x,m)=\sum\limits_{n=0}^{\infty }n^{m}\frac{x^{n}}{n!}%
\sum\limits_{k=0}^{\infty }\frac{(k+n)!}{k!}\frac{(-x)^{k}}{\Gamma (\mu
(k+n)+1)}=
\end{equation*}

\begin{equation}
\sum\limits_{n=0}^{\infty }n^{m}\frac{x^{n}}{n!}\sum\limits_{l=0}^{\infty
}\theta (l-n)\frac{l!}{(l-n)!}\frac{(-x)^{l-n}}{\Gamma (\mu l+1)},
\label{eq57}
\end{equation}

here $\theta (l)$ is the Heaviside step function,

\begin{equation}
\theta (l)=\mid \QATOP{1,\quad \ \mathrm{if\ }l\geq 0}{0,\ \quad \mathrm{if\ 
}l<0}.  \label{eq58}
\end{equation}

Then, interchanging the order of summations in Eq.(\ref{eq57}) yields

\begin{equation}
B_{\mu }(x,m)=\sum\limits_{l=0}^{\infty }\frac{x^{l}}{\Gamma (\mu l+1)}%
\sum\limits_{n=0}^{l}n^{m}\frac{(-1)^{l-n}l!}{n!(l-n)!}=\sum\limits_{l=0}^{%
\infty }\frac{x^{l}}{\Gamma (\mu l+1)}\sum\limits_{n=0}^{l}(-1)^{l-n}\binom{l%
}{n}n^{m},  \label{eq59}
\end{equation}

where the notation$\binom ln=\frac{l!}{n!(l-n)!}$ has been introduced.

By comparing Eq.(\ref{eq55}) and Eq.(\ref{eq59}) we conclude that the
fractional Stirling numbers $S_{\mu }(m,l)$ are given by 
\begin{equation}
S_{\mu }(m,l)=\frac{1}{\Gamma (\mu l+1)}\sum\limits_{n=0}^{l}(-1)^{l-n}%
\binom{l}{n}n^{m},  \label{eq60}
\end{equation}

\begin{equation*}
S_{\mu }(m,0)=\delta _{m,0},\qquad S_{\mu }(m,l)=0,\quad l=m+1,\quad m+2,....
\end{equation*}

As an example, Table 4 presents a few of fractional Stirling numbers of the
second kind.

\begin{tabular}{|l|l|l|l|l|l|l|l|}
\hline
m\textit{\TEXTsymbol{\backslash}l} & $1$ & $2$ & $3$ & $4$ & $5$ & $6$ & $7$
\\ \hline
1 & $\frac{1}{\Gamma (\mu +1)}$ &  &  &  &  &  &  \\ \hline
2 & $\frac{1}{\Gamma (\mu +1)}$ & $\frac{2}{\Gamma (2\mu +1)}$ &  &  &  &  & 
\\ \hline
3 & $\frac{1}{\Gamma (\mu +1)}$ & $\frac{6}{\Gamma (2\mu +1)}$ & $\frac{6}{%
\Gamma (3\mu +1)}$ &  &  &  &  \\ \hline
4 & $\frac{1}{\Gamma (\mu +1)}$ & $\frac{14}{\Gamma (2\mu +1)}$ & $\frac{36}{%
\Gamma (3\mu +1)}$ & $\frac{24}{\Gamma (4\mu +1)}$ &  &  &  \\ \hline
5 & $\frac{1}{\Gamma (\mu +1)}$ & $\frac{30}{\Gamma (2\mu +1)}$ & $\frac{150%
}{\Gamma (3\mu +1)}$ & $\frac{240}{\Gamma (4\mu +1)}$ & $\frac{120}{\Gamma
(5\mu +1)}$ &  &  \\ \hline
6 & $\frac{1}{\Gamma (\mu +1)}$ & $\frac{62}{\Gamma (2\mu +1)}$ & $\frac{540%
}{\Gamma (3\mu +1)}$ & $\frac{1560}{\Gamma (4\mu +1)}$ & $\frac{1800}{\Gamma
(5\mu +1)}$ & $\frac{720}{\Gamma (6\mu +1)}$ &  \\ \hline
7 & $\frac{1}{\Gamma (\mu +1)}$ & $\frac{126}{\Gamma (2\mu +1)}$ & $\frac{%
1806}{\Gamma (3\mu +1)}$ & $\frac{8400}{\Gamma (4\mu +1)}$ & $\frac{16800}{%
\Gamma (5\mu +1)}$ & $\frac{15120}{\Gamma (6\mu +1)}$ & $\frac{5040}{\Gamma
(7\mu +1)}$ \\ \hline
\end{tabular}

Table 4. \textit{Fractional Stirling numbers of the second kind }$S_{\mu
}(m,l)$\textit{\ }($0<\mu \leq 1$).

Some special cases are

\begin{equation}
S_\mu (m,1)=\frac 1{\Gamma (\mu +1)},\qquad S_\mu (m,2)=(2^{m-1}-1)\frac
2{\Gamma (2\mu +1)},  \label{eq61}
\end{equation}

\begin{equation}
S_\mu (m,3)=(3^m-3\cdot 2^m+3)\frac 1{\Gamma (3\mu +1)},  \label{eq62}
\end{equation}

\begin{equation}
S_\mu (m,4)=(4^m-4\cdot 3^m+6\cdot 2^m-4)\frac 1{\Gamma (4\mu +1)},
\label{eq63}
\end{equation}

\begin{equation}
S_\mu (m,m-1)=\frac{m!\cdot (m-1)}{2\Gamma ((m-1)\mu +1)},\qquad S_\mu (m,m)=%
\frac{m!}{\Gamma (m\mu +1)}.  \label{eq64}
\end{equation}

It is easy to see that at $\mu =1$ Eq.(\ref{eq60}) turns into the well known
representation for the standard Stirling numbers $S(m,l)=S_{\mu }(m,l)|_{\mu
=1}\equiv S_{1}(m,l)$ of the second kind \cite{Abramowitz},

\begin{equation*}
S(m,l)=\frac 1{l!}\sum\limits_{n=0}^l(-1)^{l-n}\binom lnn^m.
\end{equation*}

Thus, one can conclude that there is a relationship between fractional
Stirling numbers $S_{\mu }(m,l)$ of the second kind and standard Stirling
numbers $S(m,l)$ of the second kind

\begin{equation}
S_{\mu }(m,l)=\frac{l!}{\Gamma (\mu l+1)}S(m,l).  \label{eq65}
\end{equation}

or

\begin{equation}
S(m,l)=\frac{\Gamma (\mu l+1)}{l!}S_{\mu }(m,l).  \label{eq66}
\end{equation}

Let's note that Eqs.(\ref{eq65}) or (\ref{eq66}) allow us to find new
equations and identities for the fractional Stirling numbers $S_{\mu }(m,l)$
based on the well-know equations and identities for the standard Stirling
numbers $S(m,l)$ of the second kind$.$ For example, considering the
recurrence relation for Stirling numbers of the second kind \cite{Abramowitz}%
, (see, page 825)

\begin{equation*}
S(m+1,l)=lS(m,l)+S(m,l-1),
\end{equation*}

and using Eq.(\ref{eq65}) yield the new recurrence relation for fractional
Stirling numbers $S_{\mu }(m,l)$ of the second kind

\begin{equation}
S_{\mu }(m+1,l)=lS_{\mu }(m,l)+l\frac{\Gamma (\mu (l-1)+1)}{\Gamma (\mu l+1)}%
S_{\mu }(m,l-1).  \label{eq66a}
\end{equation}

To find a generating function of the fractional Stirling numbers $S_{\mu
}(m,l)$ of the second kind, let's expand the generating function $F_{\mu
}(s,x)$ given by Eq.(\ref{eq42}). Upon substituting $B_{\mu }(x,m)$ from Eq.(%
\ref{eq55}) we have the following chain of transformations

\begin{equation*}
F_{\mu }(s,x)=\sum\limits_{m=0}^{\infty }\frac{s^{m}}{m!}\left(
\sum\limits_{l=0}^{m}S_{\mu }(m,l)x^{l}\right) =
\end{equation*}

\begin{equation}
\sum\limits_{m=0}^\infty \frac{s^m}{m!}\left( \sum\limits_{l=0}^\infty
\theta (m-l)S_\mu (m,l)x^l\right) =\sum\limits_{l=0}^\infty \left(
\sum\limits_{m=l}^\infty S_\mu (m,l)\frac{s^m}{m!}\right) x^l,  \label{eq67}
\end{equation}

where $\theta (m-l)$ is the Heaviside step function defined by Eq.(\ref{eq58}%
).

On the other hand, from Eq.(\ref{eq45}), we have for $F_{\mu }(s,x)$

\begin{equation}
F_{\mu }(s,x)=\sum\limits_{l=0}^{\infty }\frac{(e^{s}-1)^{l}}{\Gamma (\mu
l+1)}x^{l}.  \label{eq68}
\end{equation}

Upon comparing this equation and Eq.(\ref{eq67}), we conclude that

\begin{equation}
\sum\limits_{m=l}^{\infty }S_{\mu }(m,l)\frac{s^{m}}{m!}=\frac{(e^{s}-1)^{l}%
}{\Gamma (\mu l+1)},\qquad l=0,1,2,....  \label{eq69}
\end{equation}

Now we are set up to introduce two generating functions $\mathcal{G}_{\mu
}(s,l)$ and $\mathcal{F}_{\mu }(s,t)$ of the fractional Stirling numbers of
the second kind,

\begin{equation}
\mathcal{G}_{\mu }(s,l)=\sum\limits_{m=l}^{\infty }S_{\mu }(m,l)\frac{s^{m}}{%
m!}=\frac{(e^{s}-1)^{l}}{\Gamma (\mu l+1)},  \label{eq70}
\end{equation}

\begin{equation}
\mathcal{F}_{\mu }(s,t)=\sum\limits_{m=0}^{\infty
}\sum\limits_{l=0}^{m}S_{\mu }(m,l)\frac{s^{m}t^{l}}{m!}=\sum\limits_{l=0}^{%
\infty }\frac{t^{l}(e^{s}-1)^{l}}{\Gamma (\mu l+1)}=E_{\mu }(t(e^{s}-1)).
\label{eq71}
\end{equation}

Hence, it follows from Eq.(\ref{eq70}) that

\begin{equation}
S_{\mu }(m,l)=\frac{\partial ^{m}\mathcal{G}_{\mu }(s,l)}{\partial s^{m}}%
|_{s=0}=\frac{1}{\Gamma (\mu l+1)}\frac{\partial ^{m}}{\partial s^{m}}%
(e^{s}-1)^{l}|_{s=0},  \label{eq70.1}
\end{equation}

and from Eq.(\ref{eq71}) we have for $S_{\mu }(m,l)$%
\begin{equation}
S_{\mu }(m,l)=\frac{1}{l!}\frac{\partial ^{m+l}\mathcal{F}_{\mu }(s,t)}{%
\partial s^{m}\partial t^{l}}|_{s=0,t=0}=\frac{1}{l!}\frac{\partial ^{m+l}}{%
\partial s^{m}\partial t^{l}}E_{\mu }(t(e^{s}-1))|_{s=0,t=0},\qquad l\leq m.
\label{eq71.1}
\end{equation}

As a special case $\mu =1$, equations (\ref{eq70}) and (\ref{eq71}) include
the well-know generating function equations for the standard Stirling
numbers of the second kind $S(m,l)$ (for instance, see Eqs.(2.17) and (2.18)
in Ref.\cite{Charalambides}),

\begin{equation}
\mathcal{G}_{1}(s,l)=\mathcal{G}_{\mu }(s,l)\mid _{\mu
=1}=\sum\limits_{m=l}^{\infty }S(m,l)\frac{s^{m}}{m!}=\frac{(e^{s}-1)^{l}}{l!%
},\qquad l=0,1,2,....  \label{eq72}
\end{equation}

and

\begin{equation}
\mathcal{F}_{1}(s,t)=\mathcal{F}_{\mu }(s,t)\mid _{\mu
=1}=\sum\limits_{m=0}^{\infty }\sum\limits_{l=0}^{m}S(m,l)\frac{s^{m}t^{l}}{%
m!}=\exp (t(e^{s}-1)).  \label{eq73}
\end{equation}

To get some insight on where fractional Stirling numbers of the second kind
may come from, let us prove the following lemma.

\textit{Lemma:}

If the function $A_{\mu }(s)$ can be presented by the series expansion

\begin{equation}
A_{\mu }(s)=\sum\limits_{l=0}^{\infty }a_{l}\frac{s^{l}}{\Gamma (\mu l+1)}%
,\qquad 0<\mu \leq 1,  \label{eq73.1}
\end{equation}

then

\begin{equation}
A_{\mu }(e^{s}-1)=\sum\limits_{m=0}^{\infty }b_{m}\frac{s^{m}}{m!},\qquad
0<\mu \leq 1.  \label{eq73.2}
\end{equation}

where numbers $b_{m}$ are related to the numbers $a_{l}$ by means of the
relationship

\begin{equation}
b_{m}=\sum\limits_{l=0}^{m}S_{\mu }(m,l)a_{l},  \label{eq73.3}
\end{equation}

with $S_{\mu }(m,l)$ are being fractional Stirling numbers of the second
kind introduced by Eq.(\ref{eq60}).

\textit{Proof.}

Upon substituting $e^{s}-1$ instead of $s$ into Eq.(\ref{eq73.1}) we have
the following chain of transformations

\begin{equation*}
A_{\mu }(e^{s}-1)=\sum\limits_{l=0}^{\infty }a_{l}\frac{(e^{s}-1)^{l}}{%
\Gamma (\mu l+1)}=\sum\limits_{l=0}^{\infty }\frac{a_{l}}{\Gamma (\mu l+1)}%
\sum\limits_{k=0}^{l}(-1)^{l-k}\binom{l}{k}\sum\limits_{m=0}^{\infty }\frac{%
(sk)^{m}}{m!}=
\end{equation*}

\begin{equation*}
\sum\limits_{m=0}^{\infty }\frac{s^{m}}{m!}\sum\limits_{l=0}^{m}\frac{a_{l}}{%
\Gamma (\mu l+1)}\sum\limits_{k=0}^{l}(-1)^{l-k}\binom{l}{k}%
k^{m}=\sum\limits_{n=0}^{\infty }b_{m}\frac{s^{m}}{m!},
\end{equation*}

where $b_{m}$ is given by Eq.(\ref{eq73.3}), with $S_{\mu }(m,l)$ defined by
Eq.(\ref{eq60}) and the condition $S_{\mu }(m,l)=0,\quad l\geq m+1$ has been
taken into account.

Thus, we proved the lemma.

Table 5 summarizes equations for the Bell polynomials $B(x,m)$, the Bell
numbers $B(m)$, the Stirling numbers of the second kind $S(m,l)$, the $m$-th
order moment $\overline{n^{m}}$, generating function of the Stirling numbers
of the second kind $\sum\limits_{m=l}^{\infty }S_{\mu }(m,l)s^{m}/m!$, and
generating function $\mathcal{B}(s)$ of the Bell numbers, attributed to the
fractional Poisson distribution with those for the standard Poisson
distribution.

\begin{tabular}{|c|c|c|}
\hline
& fractional Poisson ($0<\mu \leq 1)$ & Poisson ($\mu =1)$ \\ \hline
$B(x,m)$ & $\sum\limits_{n=0}^{\infty }n^{m}\frac{x^{n}}{n!}%
\sum\limits_{k=0}^{\infty }\frac{(k+n)!}{k!}\frac{(-x)^{k}}{\Gamma (\mu
(k+n)+1)}$ & $e^{-x}\sum\limits_{n=0}^{\infty }n^{m}\frac{x^{n}}{n!}$ \\ 
\hline
$B(m)$ & $\sum\limits_{n=0}^{\infty }\frac{n^{m}}{n!}E_{\mu }^{(n)}(-1)$ & $%
e^{-x}\sum\limits_{n=0}^{\infty }\frac{n^{m}}{n!}$ \\ \hline
$S(m,l)$ & $\frac{1}{\Gamma (\mu l+1)}\sum\limits_{n=0}^{l}(-1)^{l-n}\binom{l%
}{n}n^{m}$ & $\frac{1}{l!}\sum\limits_{n=0}^{l}(-1)^{l-n}\binom{l}{n}n^{m}$
\\ \hline
$\overline{n^{m}}$ & $\sum\limits_{l=0}^{m}S_{\mu }(m,l)(\nu t^{\mu })^{l}$
& $\sum\limits_{l=0}^{m}S(m,l)(\overline{\nu }t)^{l}$ \\ \hline
$\sum\limits_{m=l}^{\infty }S_{\mu }(m,l)\frac{s^{m}}{m!}$ & $\frac{%
(e^{s}-1)^{l}}{\Gamma (\mu l+1)}$ & $\frac{(e^{s}-1)^{l}}{l!}$ \\ \hline
$\mathcal{B}(s)$ & $E_{\mu }(e^{s}-1)$ & $\exp (e^{s}-1)$ \\ \hline
\end{tabular}

Table 5. \textit{Polynomials, numbers, moments and generating functions
attributed to the fractional Poisson process vs the standard Poisson process}

\subsubsection{A new representation for the Mittag-Leffler function}

To obtain a new representation for the Mittag-Leffler function defined by
Eq.(\ref{eq6}) we use the generating function of the Stirling numbers of the
first kind\footnote{%
The Stirling numbers of the first kind are defined as the coefficients $%
s(m,l)$ in the expansion
\par
\begin{equation}
\frac{x!}{(x-m)!}=x(x-1)...(x-m+1)=\sum\limits_{l=0}^{m}s(m,l)x^{l}.
\label{eqML1.1}
\end{equation}%
\par
{}}

\begin{equation}
(1+t)^{n}=\sum\limits_{m=0}^{\infty }\frac{t^{m}}{m!}\sum%
\limits_{l=0}^{m}s(m,l)n^{l},  \label{eqML1}
\end{equation}

where $s(m,l)$ stands for the Stirling numbers of the first kind \cite%
{Abramowitz2}.

Then, we evaluate expectations of both sides of Eq.(\ref{eqML1})

\begin{equation}
\sum\limits_{n=0}^{\infty }P_{\mu }(n,\lambda
)(1+t)^{n}=\sum\limits_{n=0}^{\infty }P_{\mu }(n,\lambda
)\sum\limits_{m=0}^{\infty }\frac{t^{m}}{m!}\sum\limits_{l=0}^{m}s(m,l)n^{l},
\label{eqML2}
\end{equation}

with fractional Poisson probability distribution $P_{\mu }(n,\lambda )$

\begin{equation}
P_{\mu }(n,\lambda )=\frac{(\lambda )^{n}}{n!}\sum\limits_{k=0}^{\infty }%
\frac{(k+n)!}{k!}\frac{(-\lambda )^{k}}{\Gamma (\mu (k+n)+1)},\qquad 0<\mu
\leq 1.  \label{eqML3}
\end{equation}

Calculation on a left side of Eq.(\ref{eqML2}) results in $E_{\mu }(\lambda
t)$ and we have

\begin{equation}
E_{\mu }(\lambda t)=\sum\limits_{n=0}^{\infty }P_{\mu }(n,\lambda
)\sum\limits_{m=0}^{\infty }\frac{t^{m}}{m!}\sum\limits_{l=0}^{m}s(m,l)n^{l}.
\label{eqML4}
\end{equation}

Therefore, we come to the new representation for the Mittag-Leffler function,

\begin{equation}
E_{\mu }(\lambda t)=\sum\limits_{m=0}^{\infty }\frac{t^{m}}{m!}%
\sum\limits_{l=0}^{m}s(m,l)B_{\mu }(\lambda ,l),  \label{eqML5}
\end{equation}

where $E_{\mu }(\lambda t)$ is the Mittag-Leffler function, $s(m,l)$ are the
Stirling numbers of the first kind and $B_{\mu }(\lambda ,l)$ are fractional
Bell polynomials introduced by Eq.(\ref{eq37}).

It is easy to see, that Eq.(\ref{eqML5}) can be written as

\begin{equation}
E_{\mu }(\lambda t)=\sum\limits_{m=0}^{\infty }\frac{t^{m}}{m!}%
\sum\limits_{l=0}^{m}s(m,l)\sum\limits_{r=0}^{l}S_{\mu }(l,r)\lambda ^{r},
\label{eqML6}
\end{equation}

where the representation (\ref{eq55}) of fractional Bell polynomials in
terms of fractional Stirling numbers of the second kind $S_{\mu }(l,r)$ has
been used. The last equation can be written as

\begin{equation}
E_{\mu }(\lambda t)=\sum\limits_{m=0}^{\infty }c_{m}(\mu ,\lambda )\frac{%
t^{m}}{m!},  \label{eqML6_1}
\end{equation}

if we introduce coefficients $c_{m}(\mu ,\lambda )$

\begin{equation}
c_{m}(\mu ,\lambda )=\sum\limits_{l=0}^{m}s(m,l)B_{\mu }(\lambda
,l)=\sum\limits_{l=0}^{m}s(m,l)\sum\limits_{r=0}^{l}S_{\mu }(l,r)\lambda
^{r}.  \label{eqML_2}
\end{equation}

We see from Eq.(\ref{eqML5}) that the Mittag-Leffler function $E_{\mu
}(\lambda t)$ can be considered as generating function to evaluate the sum $%
\sum\limits_{l=0}^{m}s(m,l)B_{\mu }(\lambda ,l)$, that is

\begin{equation}
\sum\limits_{l=0}^{m}s(m,l)B_{\mu }(\lambda ,l)=\frac{\partial ^{m}}{%
\partial t^{m}}E_{\mu }(\lambda t)|_{t=0},  \label{eqML6.1}
\end{equation}

or

\begin{equation}
\sum\limits_{l=0}^{m}s(m,l)\sum\limits_{r=0}^{l}S_{\mu }(l,r)\lambda ^{r}=%
\frac{\partial ^{m}}{\partial t^{m}}E_{\mu }(\lambda t)|_{t=0}.
\label{eqML6.1.1}
\end{equation}

The new formulas (\ref{eqML5})-(\ref{eqML6_1}) present the Mittag-Leffler
function $E_{\mu }(\lambda t)$ in terms of the Stirling numbers of the first
kind $s(m,l)$ and either fractional Bell polynomials $B_{\mu }(\lambda ,l)$
or fractional Stirling numbers of the second kind $S_{\mu }(l,r)$.

Comparing Eq.(\ref{eq6}) and Eq.(\ref{eqML5}) yields two new identities

\begin{equation}
\sum\limits_{l=0}^{m}s(m,l)B_{\mu }(\lambda ,l)=\frac{m!}{\Gamma (\mu m+1)}%
\lambda ^{m},  \label{eqML7}
\end{equation}

and

\begin{equation}
\sum\limits_{l=0}^{m}s(m,l)\sum\limits_{r=0}^{l}S_{\mu }(l,r)\lambda ^{r}=%
\frac{m!}{\Gamma (\mu m+1)}\lambda ^{m},  \label{eqML8}
\end{equation}

where Eq.(\ref{eq55}) has been taken into account.

At the limit case $\mu =1$ we have from these two equations

\begin{equation}
\sum\limits_{l=0}^{m}s(m,l)B(\lambda ,l)=\lambda ^{m},  \label{eqML9}
\end{equation}

or

\begin{equation}
\sum\limits_{l=0}^{m}s(m,l)\sum\limits_{r=0}^{l}S(l,r)\lambda ^{r}=\lambda
^{m}.  \label{eqML10}
\end{equation}

Hence, Eqs.(\ref{eqML5})-(\ref{eqML6_1}) are transformed into the series
expansion for the exponential function

\begin{equation*}
e^{\lambda t}=\sum\limits_{m=0}^{\infty }\frac{t^{m}}{m!}\lambda ^{m},
\end{equation*}

and coefficients $c_{m}(\mu ,\lambda )$ at $\mu =1$ become $c_{m}(\mu
,\lambda )|_{\mu =1}=c_{m}(1,\lambda )=\lambda ^{m}$.

Finally, let us note that Eq.(\ref{eqML8}) can be proved straightforwardly.
Indeed, if we substitute

\begin{equation*}
S_{\mu }(l,r)=\frac{r!}{\Gamma (\mu r+1)}S(l,r),
\end{equation*}

into Eq.(\ref{eqML8}), where $S(l,r)$ are the standard Stirling numbers of
the second kind, then we have the following chain of transformations

\begin{equation*}
\sum\limits_{l=0}^{m}s(m,l)\sum\limits_{r=0}^{l}\frac{r!}{\Gamma (\mu r+1)}%
S(l,r)\lambda ^{r}=\sum\limits_{l=0}^{m}s(m,l)\sum\limits_{r=0}^{m}\theta
(l-r)\frac{r!}{\Gamma (\mu r+1)}S(l,r)\lambda ^{r}=
\end{equation*}

\begin{equation*}
\sum\limits_{r=0}^{\infty }\frac{\lambda ^{r}r!}{\Gamma (\mu r+1)}%
\sum\limits_{l=r}^{m}s(m,l)S(l,r)=\frac{m!}{\Gamma (\mu m+1)}\lambda ^{m}.
\end{equation*}

At the last step the equation \cite{Abramowitz}

\begin{equation*}
\sum\limits_{l=r}^{m}s(m,l)S(l,r)=\delta _{m,r},
\end{equation*}

has been used, and $\delta _{m,r}$ is the Kronecker symbol.

Thus, we proved Eq.(\ref{eqML8}).

\subsubsection{The Bernoulli numbers and fractional Stirling numbers of the
second kind}

The Bernoulli numbers $B_{n}$, $n=0,1,2,...$have the generating function 
\cite{Abramowitz1},

\begin{equation}
\sum\limits_{n=0}^{\infty }B_{n}\frac{t^{n}}{n!}=\frac{t}{e^{t}-1}.
\label{eq74}
\end{equation}

These numbers play an important role in the number theory.

Let's show that the Bernoulli numbers $B_{n}$ can be presented in terms of
the fractional Stirling numbers of the second kind $S_{\mu }(n,k)$ in the
following way

\begin{equation}
B_{n}=\sum\limits_{k=0}^{n}(-1)^{k}\Gamma (\mu k+1)\frac{S_{\mu }(n,k)}{k+1}.
\label{eq75}
\end{equation}

To prove Eq.(\ref{eq75}) we substitute $B_n$ from Eq.(\ref{eq75}) into the
left-hand side of Eq.(\ref{eq74}). Therefore, we have

\begin{equation}
\sum\limits_{n=0}^{\infty }B_{n}\frac{t^{n}}{n!}=\sum\limits_{n=0}^{\infty
}\sum\limits_{k=0}^{n}(-1)^{k}\Gamma (\mu k+1)\frac{S_{\mu }(n,k)}{k+1}\frac{%
t^{n}}{n!}=  \label{eq76}
\end{equation}

\begin{equation*}
\sum\limits_{k=0}^{\infty }\sum\limits_{n=k}^{\infty }(-1)^{k}\Gamma (\mu
k+1)\frac{S_{\mu }(n,k)}{k+1}\frac{t^{n}}{n!}.
\end{equation*}

To transform the right-hand side of Eq.(\ref{eq76}) we use Eq.(\ref{eq70})
and obtain

\begin{equation*}
\sum\limits_{n=0}^\infty B_n\frac{t^n}{n!}=\sum\limits_{k=0}^\infty (-1)^k%
\frac{(e^t-1)^k}{k+1}=\frac t{e^t-1}.
\end{equation*}

Thus, we have proved Eq.(\ref{eq75}).

In the case $\mu =1$, when the gamma function is $\Gamma (\mu k+1)|_{\mu
=1}=k!$, Eq.(\ref{eq75}) reads

\begin{equation}
B_{n}=\sum\limits_{k=0}^{n}(-1)^{k}k!\frac{S(n,k)}{k+1},  \label{eq77}
\end{equation}

and we recover the representation of the Bernoulli numbers $B_{n}$ in terms
of the Stirling numbers of the second kind $S(n,k)$ (for instance, see the
equation for $B_{n}$ on page 2547 of Ref.\cite{Charalambides}).

\subsubsection{The Schl\"{a}fli polynomials and fractional Stirling numbers
of the second kind}

In 1858 L. Schl\"{a}fli (see page 31 in \cite{Schlafli}), pointed out that
the numbers $A_{n}$ introduced by means

\begin{equation}
A_{n}=\sum\limits_{k=0}^{n-1}\binom{n}{k}A_{k},  \label{eqSc1}
\end{equation}

can be obtained from the generating function

\begin{equation}
\frac{1}{2-e^{t}}=\sum\limits_{n=0}^{\infty }A_{n}\frac{t^{n}}{n!}.
\label{eqSc2}
\end{equation}

We will call the numbers $A_{n}$ as the Schl\"{a}fli numbers. The numbers $%
A_{n}$ can be expressed as

\begin{equation}
A_{n}=\sum\limits_{k=0}^{n}S(n,k)k!,  \label{eqSc3}
\end{equation}

where $S(n,k)$ are the Stirling numbers of the second kind.

Now we introduce the Schl\"{a}fli polynomials $A_{n}(x)$ defined by

\begin{equation}
A_{n}(x)=\sum\limits_{k=0}^{n}S(n,k)k!x^{k}.  \label{eqF1}
\end{equation}

As an example, here are a few Schl\"{a}fli polynomials

\begin{equation*}
A_{0}(x)=1,\qquad A_{1}(x)=x,\qquad A_{2}(x)=2x^{2}+x,\qquad ...\quad .
\end{equation*}

Taking into account Eq.(\ref{eq66}) we can express $A_{n}(x)$ as

\begin{equation}
A_{n}(x)=\sum\limits_{k=0}^{n}S_{\mu }(n,k)\Gamma (\mu k+1)x^{k},
\label{eqF2}
\end{equation}

and

\begin{equation}
A_{n}=\sum\limits_{k=0}^{n}S_{\mu }(n,k)\Gamma (\mu k+1),  \label{eqF2.2}
\end{equation}

where $S_{\mu }(n,k)$ are the fractional Stirling numbers of the second
kind. Thus, we found the new representations of the Schl\"{a}fli polynomials 
$A_{n}(x)$ and the Schl\"{a}fli numbers $A_{n}$ in terms of the fractional
Stirling numbers of the second kind.

The generating function $\mathcal{A}(t,x)$ of the Schl\"{a}fli polynomials
introduced by

\begin{equation}
\mathcal{A}(t,x)=\sum\limits_{n=0}^{\infty }\frac{t^{n}}{n!}A_{n}(x),
\label{eqF3}
\end{equation}

can be presented in terms of fractional Stirling numbers of the second kind,

\begin{equation}
\mathcal{A}(t,x)=\sum\limits_{n=0}^{\infty }\frac{t^{n}}{n!}%
\sum\limits_{k=0}^{n}S_{\mu }(n,k)\Gamma (\mu k+1)x^{k},  \label{eqF4}
\end{equation}

if we use the definition given by Eq.(\ref{eqF2}).

Further, we have the following chain of transformations

\begin{equation*}
\mathcal{A}(t,x)=\sum\limits_{n=0}^{\infty }\frac{t^{n}}{n!}\left(
\sum\limits_{k=0}^{n}S_{\mu }(m,k)\Gamma (\mu k+1)x^{k}\right) =
\end{equation*}

\begin{equation}
\sum\limits_{n=0}^{\infty }\frac{t^{n}}{n!}\left( \sum\limits_{k=0}^{\infty
}\theta (n-k)S_{\mu }(n,k)\Gamma (\mu k+1)x^{k}\right) =
\end{equation}

\begin{equation}
\sum\limits_{k=0}^{\infty }\Gamma (\mu k+1)x^{k}\left(
\sum\limits_{m=k}^{\infty }S_{\mu }(m,k)\frac{t^{m}}{m!}\right) ,
\label{eqF4.1}
\end{equation}

where $\theta (m-l)$ is the Heaviside step function defined by Eq.(\ref{eq58}%
). Substituting into Eq.(\ref{eqF4.1}) the internal sum over $m$ with the
right-hand side of Eq.(\ref{eq69}) yields

\begin{equation}
\mathcal{A}(t,x)=\sum\limits_{k=0}^{\infty }(e^{t}-1)^{k}x^{k}=\frac{1}{%
1-x(e^{t}-1)}.  \label{eqF5}
\end{equation}

Thus, we recover the generating function of the Schl\"{a}fli polynomials
(see, for instance, Eq.(9) in Ref.\cite{Tanny}). In the special case, when $%
x=1$ Eq.(\ref{eqF5}) turns into Eq.(\ref{eqSc2}).

At $x=-1$ it follows from Eqs.(\ref{eqF4}) and (\ref{eqF5}) that

\begin{equation}
\sum\limits_{k=0}^{n}S_{\mu }(n,k)\Gamma (\mu k+1)(-1)^{k}=(-1)^{n}.
\label{eqF7}
\end{equation}

which is a new formula for the fractional Stirling numbers $S_{\mu }(n,k)$
of the second kind.

The Schl\"{a}fli polynomials $A_{n}(x)$ and fractional Bell polynomials $%
B_{\mu }(x\lambda ,n)$, are related each other by

\begin{equation}
A_{n}(x)=\int\limits_{0}^{\infty }d\lambda \mathcal{L}_{\mu }(\lambda
)B_{\mu }(x\lambda ,n),  \label{eqF8}
\end{equation}

where the kernel $\mathcal{L}_{\mu }(\lambda )$ is

\begin{equation}
\mathcal{L}_{\mu }(\lambda )=\frac{\lambda ^{(\frac{1}{\mu }-1)}}{\mu }\exp
\left\{ -\lambda ^{1/\mu }\right\} .  \label{eqF9}
\end{equation}

Equation (\ref{eqF8}) can be verified by using Eq.(\ref{eq55}).

\subsection{Fractional Stirling numbers of the first kind}

To introduce fractional Stirling numbers of the first kind let us solve for $%
x^{l}$ in Eq.(\ref{eq55}),

\begin{equation*}
1=B_{\mu }(x,0),
\end{equation*}

\begin{equation*}
x=\Gamma (\mu +1)B_{\mu }(x,1),
\end{equation*}

\begin{equation}
x^{2}=\frac{\Gamma (2\mu +1)}{2!}\left( B_{\mu }(x,2)-B_{\mu }(x,1)\right) ,
\label{St1.1}
\end{equation}

\begin{equation*}
x^{3}=\frac{\Gamma (3\mu +1)}{3!}\left( B_{\mu }(x,3)-3B_{\mu }(x,2)+2B_{\mu
}(x,1)\right) ,
\end{equation*}

\begin{equation*}
x^{4}=\frac{\Gamma (4\mu +1)}{4!}\left( B_{\mu }(x,4)-6B_{\mu
}(x,3)+11B_{\mu }(x,2)-6B_{\mu }(x,1)\right) ,
\end{equation*}

and so forth. In general case we can write

\begin{equation}
x^{n}=\dsum_{k=0}^{n}s_{\mu }(n,k)B_{\mu }(x,k),  \label{St1.2}
\end{equation}

if we introduce a new numbers $s_{\mu }(n,k)$ defined as

\begin{equation}
s_{\mu }(n,k)=\frac{\Gamma (n\mu +1)}{n!}s(n,k),  \label{St1.3}
\end{equation}

with $s(n,k)$ being the Stirling numbers of the first kind \cite{Abramowitz2}%
. We call new numbers $s_{\mu }(n,k)$ as fractional Stirling numbers of the
first kind.

It follows from Eqs.(\ref{St1.3}) that

\begin{equation}
s(n,k)=\frac{n!}{\Gamma (n\mu +1)}s_{\mu }(n,k).  \label{St1.11}
\end{equation}

This equation allows us to find new equations and identities for the
fractional Stirling numbers of the first kind $s_{\mu }(n,k)$ based on the
well-know equations and identities for the standard Stirling numbers $s(n,k)$
of the first kind$.$ For example, considering the recurrence relation for
Stirling numbers of the first kind \cite{Abramowitz2},

\begin{equation*}
s(n+1,k)=s(n,k-1)-ns(n,k),\qquad 1\leq k\leq n,
\end{equation*}

yields the recurrence relation for the fractional Stirling numbers of the
first kind

\begin{equation}
\frac{(n+1)\Gamma (n\mu +1)}{\Gamma ((n+1)\mu +1)}s_{\mu }(n+1,k)=s_{\mu
}(n,k-1)-ns_{\mu }(n,k),\qquad 1\leq k\leq n.  \label{St1.4}
\end{equation}

It is easy to check that

\begin{equation}
\dsum_{k=0}^{n}s_{\mu }(n,k)=0,\qquad n>1,  \label{St1.5}
\end{equation}

and

\begin{equation}
\dsum_{k=1}^{n}(-1)^{n-k}s_{\mu }(n,k)=\Gamma (n\mu +1).  \label{St1.6}
\end{equation}

Some special cases are

\begin{equation}
s_{\mu }(n,0)=\delta _{n,0},\qquad \qquad s_{\mu }(n,1)=\frac{(-1)^{n-1}}{n}%
\Gamma (n\mu +1),  \label{St1.7}
\end{equation}

and

\begin{equation}
s_{\mu }(n,n-1)=-\frac{\Gamma (n\mu +1)}{2(n-2)!},\qquad \qquad s_{\mu
}(n,n)=\frac{\Gamma (n\mu +1)}{n!}.  \label{St1.8}
\end{equation}

As an example, Table 6 presents a few fractional Stirling numbers of the
first kind.

\begin{tabular}{|l|l|l|l|l|l|l|}
\hline
$n\mathit{\backslash k}$ & $1$ & $2$ & $3$ & $4$ & $5$ & $6$ \\ \hline
1 & ${\small \Gamma (\mu +1)}$ &  &  &  &  &  \\ \hline
2 & -$\frac{\Gamma (2\mu +1)}{2}$ & $\frac{\Gamma (2\mu +1)}{2}$ &  &  &  & 
\\ \hline
3 & $\frac{\Gamma (3\mu +1)}{3}$ & -$\frac{\Gamma (3\mu +1)}{2}$ & $\frac{%
\Gamma (3\mu +1)}{6}$ &  &  &  \\ \hline
4 & -$\frac{\Gamma (4\mu +1)}{4}$ & -$\frac{11\Gamma (4\mu +1)}{24}$ & -$%
\frac{\Gamma (4\mu +1)}{4}$ & $\frac{\Gamma (4\mu +1)}{24}$ &  &  \\ \hline
5 & $\frac{\Gamma (5\mu +1)}{5}$ & -$\frac{5\Gamma (5\mu +1)}{12}$ & $\frac{%
7\Gamma (5\mu +1)}{24}$ & -$\frac{\Gamma (5\mu +1)}{12}$ & $\frac{\Gamma
(5\mu +1)}{120}$ &  \\ \hline
6 & -$\frac{\Gamma (6\mu +1)}{6}$ & $\frac{137\Gamma (6\mu +1)}{360}$ & -$%
\frac{5\Gamma (6\mu +1)}{16}$ & $\frac{17\Gamma (6\mu +1{\small )}}{144}$ & -%
$\frac{\Gamma (6\mu +1)}{48}$ & $\frac{\Gamma (6\mu +1)}{720}$ \\ \hline
\end{tabular}

Table 6. \textit{Fractional Stirling numbers of the first kind }$s_{\mu
}(n,k)$\textit{\ }($0<\mu \leq 1$).

\section{Statistics of the fractional Poisson probability distribution}

\subsection{Moments of the fractional Poisson probability distribution}

Now we use the fractional Stirling numbers of the second kind introduced by
Eq.(\ref{eq60}) to get the moments and the central moments of the fractional
Poisson probability distribution given by Eq.(\ref{eq7}). Indeed, by
definition of the $m$-th order moment of the fractional Poisson probability
distribution we have

\begin{equation}
\overline{n_{\mu }^{m}}=\sum\limits_{n=0}^{\infty }n^{m}P_{\mu
}(n,t)=\sum\limits_{n=0}^{\infty }n^{m}\frac{(\nu t^{\mu })^{n}}{n!}%
\sum\limits_{k=0}^{\infty }\frac{(k+n)!}{k!}\frac{(-\nu t^{\mu })^{k}}{%
\Gamma (\mu (k+n)+1)},\ \ 0<\mu \leq 1.  \label{eq78}
\end{equation}

It is easy to see that $\overline{n_{\mu }^{m}}$ is in fact the fractional
Bell polynomial $B_{\mu }(\nu t^{\mu },m)$ introduced by Eq.(\ref{eq37}),

\begin{equation}
\overline{n_{\mu }^{m}}=B_{\mu }(\nu t^{\mu },m),\qquad m=0,1,2,...,\qquad
0<\mu \leq 1.  \label{eq78a}
\end{equation}

From other side, with the help of Eqs.(\ref{eq37}) and (\ref{eq55}) we find

\begin{equation}
\overline{n_{\mu }^{m}}=\sum\limits_{l=0}^{m}S_{\mu }(m,l)(\nu t^{\mu })^{l}.
\label{eq79}
\end{equation}

Hence, the fractional Stirling numbers $S_\mu (m,l)$ of the second kind
naturally appear in the power series over $\nu t^\mu $ for the $m$-th order
moment of the fractional Poisson probability distribution.

Using analytical expressions given by Eqs.(\ref{eq60}) and (\ref{eq79}),
let's list a few moments of the fractional Poisson probability distribution

\begin{equation}
\overline{n}_{\mu }=\sum\limits_{n=0}^{\infty }nP_{\mu }(n,t)=\frac{\nu
t^{\mu }}{\Gamma (\mu +1)}=B_{\mu }(\nu t^{\mu },1),  \label{eq80}
\end{equation}

\begin{equation}
\overline{n_{\mu }^{2}}=\sum\limits_{n=0}^{\infty }n^{2}P_{\mu }(n,t)=\frac{%
2(\nu t^{\mu })^{2}}{\Gamma (2\mu +1)}+\frac{\nu t^{\mu }}{\Gamma (\mu +1)}%
=B_{\mu }(\nu t^{\mu },2),  \label{eq81}
\end{equation}

\begin{equation}
\overline{n_{\mu }^{3}}=\sum\limits_{n=0}^{\infty }n^{3}P_{\mu }(n,t)=\frac{%
6(\nu t^{\mu })^{3}}{\Gamma (3\mu +1)}+\frac{6(\nu t^{\mu })^{2}}{\Gamma
(2\mu +1)}+\frac{\nu t^{\mu }}{\Gamma (\mu +1)}=B_{\mu }(\nu t^{\mu },3),
\label{eq82}
\end{equation}

\begin{equation}
\overline{n_{\mu }^{4}}=\sum\limits_{n=0}^{\infty }n^{4}P_{\mu }(n,t)=
\label{eq83}
\end{equation}

\begin{equation}
\frac{24(\nu t^{\mu })^{4}}{\Gamma (4\mu +1)}+\frac{36(\nu t^{\mu })^{3}}{%
\Gamma (3\mu +1)}+\frac{14(\nu t^{\mu })^{2}}{\Gamma (2\mu +1)}+\frac{\nu
t^{\mu }}{\Gamma (\mu +1)}=B_{\mu }(\nu t^{\mu },4),
\end{equation}

here $B_{\mu }(\nu t^{\mu },n)$, $n=1,2,3,4$, are fractional Bell
polynomials introduced by Eqs.(\ref{eq37}).

In terms of the power series over the first order moment $\overline{n}_{\mu
} $, the above equations (\ref{eq81}) - (\ref{eq83}) read

\begin{equation}
\overline{n_{\mu }^{2}}=\frac{2(\Gamma (\mu +1))^{2}}{\Gamma (2\mu +1)}%
\overline{n}_{\mu }^{2}+\overline{n}_{\mu },  \label{eq84}
\end{equation}

\begin{equation}
\overline{n_{\mu }^{3}}=\frac{6(\Gamma (\mu +1))^{3}}{\Gamma (3\mu +1)}%
\overline{n}_{\mu }^{3}+\frac{6(\Gamma (\mu +1))^{2}}{\Gamma (2\mu +1)}%
\overline{n}_{\mu }^{2}+\overline{n}_{\mu },  \label{eq85}
\end{equation}

\begin{equation}
\overline{n_{\mu }^{4}}=\frac{24(\Gamma (\mu +1))^{4}}{\Gamma (4\mu +1)}%
\overline{n}_{\mu }^{4}+\frac{36(\Gamma (\mu +1))^{3}}{\Gamma (3\mu +1)}%
\overline{n}_{\mu }^{3}+\frac{14(\Gamma (\mu +1))^{2}}{\Gamma (2\mu +1)}%
\overline{n}_{\mu }^{2}+\overline{n}_{\mu }.  \label{eq86}
\end{equation}

For the first time the mean $\overline{n}_{\mu }$ Eq.(\ref{eq80}) and the
second order moment $\overline{n_{\mu }^{2}}$ were obtained by Laskin%
\footnote{%
The second order moment defined by Eq.(\ref{eq84}) can be presented as
Eq.(27) of Ref.\cite{Laskin7} 
\begin{equation*}
\overline{n_{\mu }^{2}}=\sum\limits_{n=0}^{\infty }n^{2}P_{\mu }(n,t)=%
\overline{n}_{\mu }+\overline{n}_{\mu }^{2}\frac{\sqrt{\pi }\Gamma (\mu +1)}{%
2^{2\mu -1}\Gamma (\mu +\frac{1}{2})}.
\end{equation*}%
\par
if we take into account the well-know equations for the gamma function $%
\Gamma (\mu )$%
\par
\begin{equation*}
\Gamma (\mu +1)=\mu \Gamma (\mu ),\qquad \Gamma (2\mu )=\frac{2^{2\mu -1}}{%
\sqrt{\pi }}\Gamma (\mu )\cdot \Gamma (\mu +\frac{1}{2}).
\end{equation*}%
} (see, Eqs.(26) and (27) in Ref.\cite{Laskin7}).

In the case when $\mu =1$, equations (\ref{eq84}) - (\ref{eq86}) become the
well-know equations for moments of the standard Poisson probability
distribution with the parameter $\overline{n}\equiv \overline{n}_{1}=\nu t$
(for instance, see Eqs.(22) - (24) in Ref. \cite{Mathworld}).

\subsection{Variance, skewness and kurtosis of the fractional Poisson
probability distribution}

To find analytical expressions for variance, skewness and kurtosis of the
fractional Poisson probability distribution, let's introduce the central $m$%
-th order moment $M_{\mu }(m)$

\begin{equation*}
M_\mu (m)=\overline{(n_\mu -\overline{n}_\mu )^m}=\sum\limits_{n=0}^\infty
(n-\overline{n}_\mu )^mP_\mu (n,t)=
\end{equation*}

\begin{equation}
\sum\limits_{n=0}^{\infty }\sum\limits_{r=0}^{m}(-1)^{m-r}\binom{m}{r}n^{r}(%
\overline{n}_{\mu })^{m-r}P_{\mu }(n,t)=  \label{eq87}
\end{equation}

\begin{equation*}
\sum\limits_{r=0}^{m}(-1)^{m-r}\binom{m}{r}(\overline{n}_{\mu
})^{m-r}\sum\limits_{l=0}^{r}S_{\mu }(r,l)(\nu t^{\mu })^{l},
\end{equation*}

where $S_{\mu }(r,l)$ is given by Eq.(\ref{eq60}).

Hence, in terms of power series over the first order moment $\overline{n}%
_{\mu }$ given by Eq.(\ref{eq80}), we have

\begin{equation}
M_{\mu }(1)=0,  \label{eq88}
\end{equation}

\begin{equation}
M_{\mu }(2)=\left( \frac{2(\Gamma (\mu +1))^{2}}{\Gamma (2\mu +1)}-1\right) 
\overline{n}_{\mu }^{2}+\overline{n}_{\mu },  \label{eq89}
\end{equation}

\begin{equation}
M_{\mu }(3)=2\left( \frac{3(\Gamma (\mu +1))^{3}}{\Gamma (3\mu +1)}-\frac{%
3(\Gamma (\mu +1))^{2}}{\Gamma (2\mu +1)}+1\right) \overline{n}_{\mu }^{3}+
\label{eq90}
\end{equation}

\begin{equation*}
3\left( \frac{2(\Gamma (\mu +1))^{2}}{\Gamma (2\mu +1)}-1\right) \overline{n}%
_{\mu }^{2}+\overline{n}_{\mu },
\end{equation*}

\begin{equation}
M_{\mu }(4)=3\left( \frac{8(\Gamma (\mu +1))^{4}}{\Gamma (4\mu +1)}-\frac{%
8(\Gamma (\mu +1))^{3}}{\Gamma (3\mu +1)}+\frac{4(\Gamma (\mu +1))^{2}}{%
\Gamma (2\mu +1)}-1\right) \overline{n}_{\mu }^{4}+  \label{eq91}
\end{equation}

\begin{equation*}
6\left( \frac{6(\Gamma (\mu +1))^{3}}{\Gamma (3\mu +1)}-\frac{4(\Gamma (\mu
+1))^{2}}{\Gamma (2\mu +1)}+1\right) \overline{n}_{\mu }^{3}+2\left( \frac{%
7(\Gamma (\mu +1))^{2}}{\Gamma (2\mu +1)}-2\right) \overline{n}_{\mu }^{2}+%
\overline{n}_{\mu }.
\end{equation*}

Further, in terms of the above defined central moments $M_{\mu }(m)$, the
variance $\sigma ^{2}$, skewness $s_{\mu }$, and kurtosis $k_{\mu }$ of the
fractional Poisson probability distribution are

\begin{equation}
\sigma _{\mu }^{2}=M_{\mu }(2),  \label{eq92}
\end{equation}

\begin{equation}
{\large s}_{\mu }=\frac{M_{\mu }(3)}{M_{\mu }^{3/2}(2)},  \label{eq93}
\end{equation}

\begin{equation}
{\large k}_{\mu }=\frac{M_{\mu }(4)}{M_{\mu }^{2}(2)}-3.  \label{eq94}
\end{equation}

In the case when $\mu =1$, new equations (\ref{eq89}) - (\ref{eq91}) turn
into equations for the central moments of the standard Poisson probability
distribution with the parameter $\overline{n}\equiv \overline{n}_{1}=%
\overline{\nu }t$ (see, Eqs.(25) - (27) in Ref. \cite{Mathworld}).

Equations (\ref{eq92}) - (\ref{eq94}),\ at $\mu =1$, turn into the equations
for variance, skewness, and kurtosis of the standard Poisson probability
distribution with the parameter $\overline{n}\equiv \overline{n}_{1}=%
\overline{\nu }t$ (for instance, see Eqs.(29) - (31) in Ref. \cite{Mathworld}%
).

\section{New polynomials}

\subsection{Generating functions}

\subsubsection{Multiplicative renormalization framework}

It is well known from the theory of orthogonal polynomials \cite{Chihara}
that orthogonal polynomials of discrete variable are associated with
discrete probability distributions. Fractional Poisson probability
distribution is a new member of the family of discrete probability
distributions. Thus, based on our findings we now post the challenge to
design and develop a new system of polynomials associated with fractional
Poisson probability distribution

\begin{equation}
P_{\mu }(x,\lambda _{\mu })=\frac{(\lambda _{\mu })^{x}}{x!}%
\sum\limits_{k=0}^{\infty }\frac{(k+x)!}{k!}\frac{(-\lambda _{\mu })^{k}}{%
\Gamma (\mu (k+x)+1)},\qquad 0<\mu \leq 1,  \label{eq771P}
\end{equation}

where $x$ is discrete random variable ($x=0,1,2,$ ...) and $\lambda _{\mu }$
is parameter.

To meet the challenge we will follow the idea of a multiplicative
renormalization framework \cite{Kubo}. Thus, let's consider the function

\begin{equation}
\varphi (t,x)=(1+t)^{x},  \label{eq772P}
\end{equation}

where we treat $x$ as a discrete random variable with fractional Poisson
probability distribution, see, Eq.(\ref{eq771P}). Then, the multiplicative
renormalization $\psi _{\mu }(t,x)$ is defined by

\begin{equation}
\psi _{\mu }(t,x)=\frac{\varphi (t,x)}{<\varphi (t,x)>_{P_{\mu }}},
\label{q772P}
\end{equation}

here $<$...$>_{P_{\mu }}$ stands for expectation over $P_{\mu }(x,\lambda )$
given by Eq.(\ref{eq771P}), that is

\begin{equation}
<\varphi (t,x)>_{P_{\mu }}=\sum\limits_{x=0}^{\infty }P_{\mu }(x,\lambda
_{\mu })\varphi (t,x).  \label{eq773P}
\end{equation}

Then, evaluating the expectation $<$...$>_{P_{\mu }}$ yields

\begin{equation}
\psi _{\mu }(t,x)=\frac{\varphi (t,x)}{E_{\mu }(\lambda _{\mu }t)},
\label{eq774P}
\end{equation}

where $E_{\mu }(\lambda t)$ is the Mittag-Leffler function defined by Eq.(%
\ref{eq6}).

Function $\psi _{\mu }(t,x)$\ is the multiplicative renormalization of
function $\varphi (t,x)$ in Eq.(\ref{eq772P}) over the probabilistic measure
associated with fractional Poisson probability distribution Eq.(\ref{eq771P}%
).

Multiplicative renormalization framework \cite{Kubo} says that the
multiplicative renormalization $\psi (t,x)$ can be considered as a
generating function of new polynomials $L_{n}(x;\lambda _{\mu })$. That is

\begin{equation}
\psi _{\mu }(t,x)=\frac{(1+t)^{x}}{E_{\mu }(\lambda t)}=\sum\limits_{n=0}^{%
\infty }L_{n}(x;\lambda _{\mu })\frac{t^{n}}{n!},\qquad 0<\mu \leq 1,
\label{eq775P}
\end{equation}

where $n$-degree polynomials $L_{n}(x;\lambda _{\mu })$ of discrete variable 
$x$ depend on the value of real parameter $\lambda $ and parameter $\mu $
associated with fractional Poisson probability distribution.

At $x=0$ Eq.(\ref{eq775P}) can be considered as the definition for
generating function $\phi _{\mu }(t)=\psi _{\mu }(t,0)$ of new numbers $%
L_{n}(\lambda _{\mu })=L_{n}(0;\lambda _{\mu })$

\begin{equation}
\phi _{\mu }(t)=\frac{1}{E_{\mu }(\lambda _{\mu }t)}=\sum\limits_{n=0}^{%
\infty }L_{n}(\lambda _{\mu })\frac{t^{n}}{n!},\qquad 0<\mu \leq 1,
\label{eq775PN}
\end{equation}

To obtain explicit expressions for polynomials $L_{n}(x;\lambda _{\mu })$ we
write

\begin{equation}
\frac{1}{E_{\mu }(\lambda _{\mu }t)}=\sum\limits_{k=0}^{\infty }a_{k}(\mu )%
\frac{(\lambda _{\mu }t)^{k}}{k!},  \label{eq776P}
\end{equation}

with yet unknown coefficients $a_{k}(\mu )$ that have to be found.

With help of Eq.(\ref{eq776P}) the generating function $\psi _{\mu }(t,x)$
reads

\begin{equation}
\psi _{\mu }(t,x)=\sum\limits_{n=0}^{\infty }\binom{x}{n}t^{n}\sum%
\limits_{k=0}^{\infty }a_{k}(\mu )\frac{(\lambda _{\mu }t)^{k}}{k!}%
=\sum\limits_{n=0}^{\infty }\frac{t^{n}}{n!}\left( \sum\limits_{k=0}^{n}%
\frac{n!}{(n-k)!}\lambda _{\mu }^{n-k}\binom{x}{k}a_{n-k}(\mu )\right) ,
\label{eq776PP}
\end{equation}

where notation $\binom{x}{n}=\frac{x!}{n!(x-n)!}$ has been used.

Comparing Eqs.(\ref{eq775P}) and (\ref{eq776PP}) gives us an explicit
formula for new polynomials $L_{n}(x;\lambda _{\mu })$,

\begin{equation}
L_{n}(x;\lambda _{\mu })=\sum\limits_{k=0}^{n}\frac{n!}{(n-k)!}\lambda _{\mu
}^{n-k}\binom{x}{k}a_{n-k}(\mu ),  \label{eq776PPP}
\end{equation}

while comparing Eqs.(\ref{eq775PN}) and (\ref{eq776PP}) gives us explicit
formula for new numbers $L_{n}(\lambda _{\mu })$

\begin{equation}
L_{n}(\lambda _{\mu })=\lambda _{\mu }^{n}a_{n}(\mu ).  \label{eq776PPN}
\end{equation}

Polynomials $L_{n}(x;\lambda _{\mu })$ can be written as

\begin{equation}
L_{n}(x;\lambda _{\mu })=\sum\limits_{k=0}^{n}\binom{n}{k}\lambda _{\mu
}^{n-k}a_{n-k}(\mu )\dsum\limits_{l=0}^{k}s(k,l)x^{l},  \label{eq776PPA}
\end{equation}

or

\begin{equation}
L_{n}(x;\lambda _{\mu })=\dsum\limits_{l=0}^{n}x^{l}\sum\limits_{k=l}^{n}%
\binom{n}{k}\lambda _{\mu }^{n-k}a_{n-k}(\mu )s(k,l),  \label{eq776PPB}
\end{equation}

where $s(k,l)$ are the Stirling numbers of the first kind \cite{Abramowitz2}.

Aiming to find coefficients $a_{n}(\mu )$ we rewrite Eq.(\ref{eq776P}) as

\begin{equation}
\sum\limits_{m=0}^{\infty }\frac{(\lambda _{\mu }t)^{m}}{\Gamma (\mu m+1)}%
\sum\limits_{k=0}^{\infty }a_{k}(\mu )\frac{(\lambda _{\mu }t)^{k}}{k!}=1.
\label{eq777P}
\end{equation}

Then, the Cauchy product rule yields for the left hand side of Eq.(\ref%
{eq777P})

\begin{equation}
\sum\limits_{m=0}^{\infty }\frac{(\lambda _{\mu }t)^{m}}{\Gamma (\mu m+1)}%
\sum\limits_{k=0}^{\infty }a_{k}(\mu )\frac{(\lambda _{\mu }t)^{k}}{k!}%
=\sum\limits_{n=0}^{\infty }c_{n}(\mu )\frac{(\lambda _{\mu }t)^{n}}{n!},
\label{778P}
\end{equation}

where

\begin{equation}
c_{n}(\mu )=\sum\limits_{l=0}^{n}\frac{n!}{(n-l)!}\frac{a_{n-k}(\mu )}{%
\Gamma (\mu l+1)}.  \label{eq779P}
\end{equation}

From Eqs.(\ref{eq777P})-(\ref{eq779P}) we come to the conclusion that

\begin{equation}
c_{0}(\mu )=1,\qquad c_{n}(\mu )=0,\qquad n\geq 1.
\end{equation}

Hence, we have

\begin{equation}
a_{0}(\mu )=c_{0}(\mu )=1,  \label{eq780P}
\end{equation}

and

\begin{equation}
\sum\limits_{l=0}^{n}\frac{n!}{(n-l)!}\frac{a_{n-l}(\mu )}{\Gamma (\mu l+1)}%
=0,\qquad n\geq 1.  \label{eq781P}
\end{equation}

The last equation results in

\begin{equation}
a_{n}(\mu )=-\sum\limits_{l=1}^{n}\frac{n!}{(n-l)!}\frac{a_{n-l}(\mu )}{%
\Gamma (\mu l+1)},\qquad n\geq 1.  \label{eq781PP}
\end{equation}

The system of two equations (\ref{eq781PP}) and (\ref{eq780P}) can be
iterated to obtain $a_{n}(\mu ).$ As an example, here are explicit
expressions for a few coefficients $a_{n}(\mu )$

\begin{equation}
a_{1}(\mu )=-\frac{1}{\Gamma (\mu +1)},  \label{eq782P}
\end{equation}

\begin{equation}
a_{2}(\mu )=\frac{2}{(\Gamma (\mu +1))^{2}}-\frac{2}{\Gamma (2\mu +1)},
\label{eq785P}
\end{equation}

\begin{equation}
a_{3}(\mu )=-\frac{6}{(\Gamma (\mu +1))^{3}}+\frac{12}{\Gamma (\mu +1)\Gamma
(2\mu +1)}-\frac{6}{\Gamma (3\mu +1)}.  \label{eq786P}
\end{equation}

Now we can present a few new polynomials $L_{n}(x;\lambda _{\mu })$, $0<\mu
\leq 1$

\begin{equation}
L_{0}(x;\lambda _{\mu })=1,  \label{eq787P}
\end{equation}

\begin{equation}
L_{1}(x;\lambda _{\mu })=x-\frac{\lambda _{\mu }}{\Gamma (\mu +1)},
\label{eq788P}
\end{equation}

\begin{equation}
L_{2}(x;\lambda _{\mu })=x^{2}-x(1+\frac{2\lambda _{\mu }}{\Gamma (\mu +1)}%
)+2\lambda _{\mu }^{2}\left( \frac{1}{(\Gamma (\mu +1))^{2}}-\frac{1}{\Gamma
(2\mu +1)}\right) .  \label{eq789P}
\end{equation}

\begin{equation*}
L_{3}(x;\lambda _{\mu })=x^{3}-3x^{2}\left( 1+\frac{\lambda _{\mu }}{\Gamma
(\mu +1)}\right)
\end{equation*}

\begin{equation}
+x\left( 2+6\lambda _{\mu }^{2}\left( \frac{1}{(\Gamma (\mu +1))^{2}}-\frac{1%
}{\Gamma (2\mu +1)}\right) +\frac{3\lambda _{\mu }}{\Gamma (\mu +1)}\right)
\label{eq790P}
\end{equation}

\begin{equation}
+6\lambda _{\mu }^{3}\left( -\frac{1}{(\Gamma (\mu +1))^{2}}+\frac{2}{\Gamma
(\mu +1)\Gamma (2\mu +1)}-\frac{1}{\Gamma (3\mu +1)}\right) ,
\end{equation}

and a few new numbers $L_{n}(\lambda _{\mu })$

\begin{equation}
L_{0}(\lambda _{\mu })=1,  \label{eq791N}
\end{equation}

\begin{equation}
L_{1}(\lambda _{\mu })=-\frac{\lambda _{\mu }}{\Gamma (\mu +1)},
\label{eq792N}
\end{equation}

\begin{equation}
L_{2}(\lambda _{\mu })=2\lambda _{\mu }^{2}\left( \frac{1}{(\Gamma (\mu
+1))^{2}}-\frac{1}{\Gamma (2\mu +1)}\right) .  \label{eq793N}
\end{equation}

\begin{equation}
L_{3}(\lambda _{\mu })=6\lambda _{\mu }^{3}\left( -\frac{1}{(\Gamma (\mu
+1))^{2}}+\frac{2}{\Gamma (\mu +1)\Gamma (2\mu +1)}-\frac{1}{\Gamma (3\mu +1)%
}\right) .  \label{eq794N}
\end{equation}

Newly introduced polynomials $L_{n}(x;\lambda _{\mu })$ do not form a system
of orthogonal polynomials, because of non-Markov property of the fractional
Poisson probability distribution. To illustrate this statement let's follow
the multiplicative renormalization framework \cite{Kubo} and calculate $%
<\psi _{\mu }(s,x)\psi _{\mu }(t,x)>_{P_{\mu }}$,

\begin{equation}
<\psi _{\mu }(s,x)\psi _{\mu }(t,x)>_{P_{\mu }}=\sum\limits_{x=0}^{\infty
}P_{\mu }(x,\lambda _{\mu })\frac{(1+s)^{x}}{E_{\mu }(\lambda _{\mu }s)}%
\frac{(1+t)^{x}}{E_{\mu }(\lambda _{\mu }t)}=  \label{eq791P}
\end{equation}%
\begin{equation*}
\frac{E_{\mu }(\lambda _{\mu }s+\lambda _{\mu }t+\lambda _{\mu }st)}{E_{\mu
}(\lambda _{\mu }s)E_{\mu }(\lambda _{\mu }t)},
\end{equation*}

where $P_{\mu }(x,\lambda _{\mu })$ is given by Eq.(\ref{eq771P}).

The multiplicative renormalization framework \cite{Kubo} says that if, and
only if, the expectation of the product of two generating functions $<\psi
(t,x)\psi (s,x)>_{P_{\mu }}$ is function of $st$, then it implies
orthogonality of the polynomials generated by $\psi _{\mu }(t,x)$. It is not
the case for the generating function Eq.(\ref{eq775P}), because of the
presence of the Mittag-Leffler functions in the right side of Eq.(\ref%
{eq791P}).

In the limit case $\mu =1$ the probability distribution function $P_{\mu
}(x,\lambda _{\mu })|_{\mu =1}$ defined by Eq.(\ref{eq771P}) becomes the
well-known Poisson probability distribution $P(x,\lambda )$ with $\lambda =$ 
$\lambda _{\mu }|_{\mu =1}$,

\begin{equation}
P(x,\lambda )=\frac{(\lambda )^{x}}{x!}e^{-\lambda },  \label{eq792P}
\end{equation}

which posses the Markov property. Further, the generating function $\psi
(t,x)=\psi _{\mu }(t,x)|_{\mu =1}$ becomes

\begin{equation}
\psi (t,x)=\frac{\varphi (t,x)}{e^{\lambda t}},  \label{eq793P}
\end{equation}

with $\varphi (t,x)$ given by Eq.(\ref{eq772P}). Thus, we obtain,

\begin{equation}
\psi (t,x)=(1+t)^{x}e^{-\lambda t}=\sum\limits_{n=0}^{\infty }\frac{t^{n}}{n!%
}C_{n}(x;\lambda ),  \label{eq794P}
\end{equation}

which is recognizable as the generating function of the Charlier orthogonal
polynomials $C_{n}(x;\lambda )$ of discrete variable $x$ \cite{Chihara}
associated with the Poisson probability distribution.

It is easy to see, that

\begin{equation}
<\psi (s,x)\psi (t,x)>_{P}=\sum\limits_{x=0}^{\infty }P(x,\lambda
)(1+s)^{x}e^{-\lambda s}(1+t)^{x}e^{-\lambda t}=e^{\lambda st},
\label{eq795P}
\end{equation}

which means that $<\psi (s,x)\psi (t,x)>_{P}$ depends on $st$ only, and we
conclude that function $\psi (t,x)$ introduced by Eq.(\ref{eq794P}) is in
fact the generating function of orthogonal polynomials. Thus, in the limit
case $\mu =1$ polynomials $L_{n}(x;\lambda ,\mu )$ become the Charlier
orthogonal polynomials $C_{n}(x;\lambda )$

\begin{equation}
L_{n}(x;\lambda _{\mu })|_{\mu =1}=C_{n}(x;\lambda ),\qquad \lambda =\lambda
_{\mu }|_{\mu =1}.  \label{eq795.1P}
\end{equation}

In other words, at $\mu =1$ when the fractional Poisson probability
distribution becomes the standards Poisson distribution and Markov property
is restored, newly introduced non-orthogonal polynomials $L_{n}(x;\lambda
_{\mu })$ become the well-known orthogonal Charlier polynomials $%
C_{n}(x;\lambda ).$ Hence, the polynomials $L_{n}(x;\lambda _{\mu })$ can be
considered as a generalization of the Charlier orthogonal polynomials $%
C_{n}(x;\lambda ).$

At $\mu =1$ it follows from Eq.(\ref{eq776P}) that $e^{-\lambda
t}=\sum\limits_{k=0}^{\infty }a_{k}(\mu )\frac{(\lambda t)^{k}}{k!},$ and we
conclude that $a_{k}(\mu )=(-1)^{n}$. Hence, the generating function $\psi
(t,x)$ for the Charlier polynomials is

\begin{equation}
\psi (t,x)=\sum\limits_{n=0}^{\infty }\binom{x}{n}t^{n}\sum\limits_{k=0}^{%
\infty }\frac{(-\lambda t)^{k}}{k!}=\sum\limits_{n=0}^{\infty }\frac{t^{n}}{%
n!}\left( \sum\limits_{k=0}^{n}\frac{n!}{(n-k)!}(-\lambda )^{n-k}\binom{x}{k}%
\right) ,  \label{eq795C}
\end{equation}

and explicit expression for the Charlier polynomials $C_{n}(x;\lambda )$ is

\begin{equation}
C_{n}(x;\lambda )=\sum\limits_{k=0}^{n}\frac{n!}{(n-k)!}(-\lambda )^{n-k}%
\binom{x}{k},  \label{eq796C}
\end{equation}

or in terms of Stirling numbers of the first kind (see, Eq.(\ref{eqML1.1}))

\begin{equation}
C_{n}(x;\lambda )=\dsum\limits_{l=0}^{n}x^{l}\sum\limits_{k=l}^{n}\binom{n}{k%
}(-\lambda )^{n-k}s(k,l).  \label{eq796St}
\end{equation}

A few Charlier polynomials $C_{n}(x;\lambda )$ can be found from Eqs.(\ref%
{eq787P}) - (\ref{eq790P}) at $\mu =1$, or from Eq.(\ref{eq796C}), and they
are

\begin{equation}
C_{0}(x;\lambda )=L_{0}(x;\lambda _{\mu })|_{\mu =1}=1,  \label{eq796P}
\end{equation}

\begin{equation}
C_{1}(x;\lambda )=L_{1}(x;\lambda _{\mu })|_{\mu =1}=x-\lambda ,
\label{eq797P}
\end{equation}

\begin{equation}
C_{2}(x;\lambda )=L_{2}(x;\lambda _{\mu })|_{\mu =1}=x^{2}-x(1+2\lambda
)+\lambda ^{2},  \label{eq798P}
\end{equation}

\begin{equation}
C_{3}(x;\lambda )=L_{3}(x;\lambda _{\mu })|_{\mu =1}=x^{3}-3x^{2}(1+\lambda
)+x\left( 2+3\lambda +3\lambda ^{2}\right) -\lambda ^{3},  \label{eq799P}
\end{equation}

where $\lambda =$ $\lambda _{\mu }|_{\mu =1}$.

From Eqs.(\ref{eq794P}) and (\ref{eq795C}) we have

\begin{equation}
\sum\limits_{x=0}^{\infty }\frac{(\lambda )^{x}}{x!}e^{-\lambda
}\sum\limits_{n=0}^{\infty }\frac{s^{n}}{n!}C_{n}(x;\lambda
)\sum\limits_{m=0}^{\infty }\frac{t^{m}}{m!}C_{m}(x;\lambda
)=\sum\limits_{n=0}^{\infty }\frac{\lambda ^{n}(st)^{n}}{n!}.  \label{eq800P}
\end{equation}

Thus, we obtain

\begin{equation}
\sum\limits_{n=0}^{\infty }\sum\limits_{m=0}^{\infty }\left(
\sum\limits_{x=0}^{\infty }\frac{(\lambda )^{x}}{x!}e^{-\lambda
}C_{n}(x;\lambda )C_{m}(x;\lambda )\right) \frac{s^{n}}{n!}\frac{t^{m}}{m!}%
=\sum\limits_{n=0}^{\infty }\frac{\lambda ^{n}(st)^{n}}{n!}.  \label{eq801P}
\end{equation}

Comparison of the coefficients in Eq.(\ref{eq801P}) leads to the
orthogonality condition for the Charlier polynomials

\begin{equation}
\sum\limits_{x=0}^{\infty }\frac{(\lambda )^{x}}{x!}e^{-\lambda
}C_{n}(x;\lambda )C_{m}(x;\lambda )=\lambda ^{n}n!\delta _{n,m},
\label{eq802P}
\end{equation}

where $\delta _{n,m}$ is the Kronecker symbol.

Table 7 compares equations for newly introduced polynomials $L_{n}(x;\lambda
_{\mu })$\ vs the Charlier orthogonal polynomials $C_{n}(x;\lambda )$. Table
7 presents two sets of equations for probability distribution function $%
P(x,\lambda )$ of discrete random variable $x$, polynomials generating
function $\psi (t,x)$, generating function $\phi (t)$ of numbers,
expectation of the product of two multiplicative renormalization functions $%
<\psi (t,x)\psi (s,x)>$ and orthogonality condition.

\begin{center}
\begin{tabular}{|c|c|c|}
\hline
& {\small Polynomials }$L_{n}(x;\lambda _{\mu })${\tiny \ (}$0<\mu \leq 1)$
& {\small Charlier polynomials} $C_{n}(x;\lambda )${\tiny \ (}$\mu =1)$ \\ 
\hline
${\small P(x,\lambda )}$ & $\frac{(\lambda _{\mu })^{x}}{x!}%
\sum\limits_{k=0}^{\infty }\frac{(k+n)!}{k!}\frac{(-\lambda _{\mu })^{k}}{%
\Gamma (\mu (k+n)+1)}$ & $\frac{(\lambda )^{x}}{x!}\exp (-\lambda )$ \\ 
\hline
${\small \psi (t,x)}$ & $\frac{(1+t)^{x}}{E_{\mu }(\lambda _{\mu }t)}%
=\sum\limits_{n=0}^{\infty }L_{n}(x;\lambda _{\mu })\frac{t^{n}}{n!}$ & $%
(1+t)^{x}e^{-\lambda t}=\sum\limits_{n=0}^{\infty }\frac{t^{n}}{n!}%
C_{n}(x;\lambda )$ \\ \hline
${\small \phi (t)}$ & $\frac{1}{E_{\mu }(\lambda _{\mu }t)}%
=\sum\limits_{n=0}^{\infty }L_{n}(\lambda _{\mu })\frac{t^{n}}{n!}$ & $%
e^{-\lambda t}=\sum\limits_{n=0}^{\infty }\frac{t^{n}}{n!}C_{n}(\lambda )$
\\ \hline
${\tiny <\psi (s,x)\psi (t,x)>}$ & $\frac{E_{\mu }(\lambda _{\mu }s+\lambda
_{\mu }t+\lambda _{\mu }st)}{E_{\mu }(\lambda _{\mu }s)E_{\mu }(\lambda
_{\mu }t)}$ & $e^{\lambda st}$ \\ \hline
{\small Orthogonality} & {\small Non-orthogonal} & $\sum\limits_{x=0}^{%
\infty }\frac{(\lambda )^{x}{\small e}^{-\lambda }}{x!}{\small C}_{n}{\small %
(x;\lambda )C}_{m}{\small (x;\lambda )=\lambda }^{n}{\small n!\delta }_{n,m}$
\\ \hline
\end{tabular}
\end{center}

Table 7.\textit{\ Multiplicative renormalization formulas related to the
polynomials }$L_{n}(x;\lambda _{\mu })$\textit{\ vs the Charlier polynomials 
}$C_{n}(x;\lambda )$\textit{.}

\subsubsection{Alternative approach}

As an alternative to the multiplicative renormalization framework, we
develop an approach to build polynomial sequence $l_{n}(x;\lambda _{\mu })$
based on generating function $g_{\mu }(t,x)$ introduced by,

\begin{equation}
g_{\mu }(t,x)=(1+t)^{x}E_{\mu }(-\lambda _{\mu }t)=\sum\limits_{n=0}^{\infty
}l_{n}(x;\lambda _{\mu })\frac{t^{n}}{n!},\qquad 0<\mu \leq 1,
\label{eq771al}
\end{equation}

where $E_{\mu }(x)$ is the Mittag-Leffler function (see, definition given by
Eq.(\ref{eq6})). The $n$-degree polynomials $l_{n}(x;\lambda _{\mu })$ of
discrete variable $x$ depend on the value of real parameter $\lambda _{\mu }$
and parameter $\mu $ associated with fractional Poisson probability
distribution given by Eq.(\ref{eq771P}). The motivation behind Eq.(\ref%
{eq771al}) is an attempt to get polynomial sequence $l_{n}(x;\lambda _{\mu
}) $ which coincides with the Charlier orthogonal polynomials at $\mu =1$
and is alternative to the polynomial sequence $L_{n}(x;\lambda _{\mu })$
introduced by Eq.(\ref{eq775P}).

At $x=0$ Eq.(\ref{eq771al}) can be considered as the definition for
generating function $j_{\mu }(t)=g_{\mu }(t,0)$ of numbers $l_{n}(\lambda
_{\mu })=l_{n}(0;\lambda _{\mu })$

\begin{equation}
j_{\mu }(t)=E_{\mu }(-\lambda _{\mu }t)=\sum\limits_{n=0}^{\infty
}l_{n}(\lambda _{\mu })\frac{t^{n}}{n!},\qquad 0<\mu \leq 1.  \label{eq771an}
\end{equation}

It follows from Eq.(\ref{eq771al}) that

\begin{equation}
g_{\mu }(t,x)=\sum\limits_{n=0}^{\infty }\binom{x}{n}t^{n}\sum%
\limits_{k=0}^{\infty }\frac{(-\lambda _{\mu }t)^{k}}{\Gamma (\mu k+1)}=
\label{eq772al}
\end{equation}

\begin{equation}
\sum\limits_{n=0}^{\infty }\frac{t^{n}}{n!}\left( \sum\limits_{k=0}^{n}\frac{%
n!}{\Gamma (\mu (n-k)+1)}(-\lambda _{\mu })^{n-k}\binom{x}{k}\right) .
\end{equation}

Thus, we have

\begin{equation}
l_{n}(x;\lambda _{\mu })=\sum\limits_{k=0}^{n}\frac{n!}{\Gamma (\mu (n-k)+1)}%
(-\lambda _{\mu })^{n-k}\binom{x}{k},  \label{eq773al}
\end{equation}

or

\begin{equation}
l_{n}(x;\lambda _{\mu })=\dsum\limits_{l=0}^{n}x^{l}\sum\limits_{k=l}^{n}%
\frac{n!}{k!\Gamma (\mu (n-k)+1)}(-\lambda _{\mu })^{n-k}s(k,l),\qquad 0<\mu
\leq 1,  \label{eq774al}
\end{equation}

where $s(k,l)$ are Stirling numbers of the first kind \cite{Abramowitz2}.

A few new polynomials $l_{n}(x;\lambda _{\mu })$, $0<\mu \leq 1$ are

\begin{equation}
l_{0}(x;\lambda _{\mu })=1,  \label{eq775al}
\end{equation}

\begin{equation}
l_{1}(x;\lambda _{\mu })=x-\frac{\lambda _{\mu }}{\Gamma (\mu +1)},
\label{eq776al}
\end{equation}

\begin{equation}
l_{2}(x;\lambda _{\mu })=x^{2}-x(1+\frac{2\lambda _{\mu }}{\Gamma (\mu +1)})+%
\frac{2\lambda _{\mu }^{2}}{\Gamma (2\mu +1)},  \label{eq777al}
\end{equation}

\begin{equation}
l_{3}(x;\lambda _{\mu })=x^{3}-3x^{2}(1+\frac{\lambda _{\mu }}{\Gamma (\mu
+1)})+x\left( 2+\frac{3\lambda _{\mu }}{\Gamma (\mu +1)}+\frac{6\lambda
_{\mu }^{2}}{\Gamma (2\mu +1)}\right) -\frac{6\lambda _{\mu }^{3}}{\Gamma
(3\mu +1)}.  \label{eq778al}
\end{equation}

The numbers $l_{n}(\lambda _{\mu })$, $0<\mu \leq 1$ are given by

\begin{equation}
l_{n}(\lambda _{\mu })=(-\lambda _{\mu })^{n}\frac{n!}{\Gamma (n\mu +1)}.
\label{eq779al}
\end{equation}

Newly introduced polynomials $l_{n}(x;\lambda _{\mu })$ do not form a system
of orthogonal polynomials. In the limit case $\mu =1$ Eq.(\ref{eq771al})
goes into Eq.(\ref{eq794P}) and we come to the generating function of the
Charlier orthogonal polynomials. Hence, at $\mu =1$ newly introduced
polynomials $l_{n}(x;\lambda _{\mu })|_{\mu =1}$ become the Charlier
orthogonal polynomials, $l_{n}(x;\lambda _{\mu })|_{\mu =1}=C_{n}(x;\lambda
) $.

\section{Conclusions}

Applications of the fractional Poisson probability distribution to quantum
physics, number theory and theory of polynomials have been presented.

As a quantum physics application, a new family of quantum coherent states
has been introduced and explored to study physical phenomena where the
distribution of photon numbers is governed by the fractional Poisson
probability distribution.

As number theory applications we have discovered and developed the
fractional generalization of Bell polynomials, Bell numbers, and Stirling
numbers of the first kind and the second kind. Appearance of fractional Bell
polynomials is natural if one evaluates the diagonal matrix element of the
evolution operator in the basis of newly introduced quantum coherent states.
Fractional Stirling numbers of the second kind have been introduced and
applied to evaluate skewness and kurtosis of the fractional Poisson
probability distribution function. A new representation of the Bernoulli
numbers in terms of fractional Stirling numbers of the second kind has been
obtained. A representation of Schl\"{a}fli polynomials in terms of
fractional Stirling numbers of the second kind has been found. The integral
relationship between the Schl\"{a}fli polynomials and fractional Bell
polynomials has been obtained. A new representation of the Mittag-Leffler
function involving fractional Bell polynomials and fractional Stirling
numbers of the second kind has been discovered. Fractional Stirling numbers
of the first kind have also been introduced and studied.

Two new sequences of polynomials of discrete variable associated with
fractional Poisson probability distribution has been launched and explored.
The relationship between new polynomials and the orthogonal Charlier
polynomials has also been investigated.

In the limit case when the fractional Poisson probability distribution
becomes the Poisson probability distribution, all of the above listed
developments and implementations turn into the well-known results of the
quantum optics, the theory of combinatorial numbers and the theory of
orthogonal polynomials of discrete variable.

Table 1 compares fundamental equations associated with the fractional
Poisson probability distribution to those of the well-known ones, related to
the standard Poisson probability distribution. Tables 2, 3, 4, 5, 6 and 7
summarize our findings for the fractional Poisson probability distribution
in comparison to the well-known results attributed to the standard Poisson
probability distribution.

These findings facilitate the further exploration of long-memory impact on
quantum phenomena and initiate studies on fundamental relationships between
orthogonality of polynomials and the Markov property of underlying
probabilistic distributions involved into multiplicative renormalization.

\section{Appendix}

To obtain Eq.(\ref{eq24}) we use the Laplace transform of the Mittag-Leffler
function $E_{\mu }(-z\tau ^{\mu })$

\begin{equation*}
\int\limits_{0}^{\infty }d\tau e^{-\tau }E_{\mu }(-z\tau ^{\mu })=\frac{1}{%
1+z},
\end{equation*}

for instance, see equation (26) on page 210 of Ref.\cite{Erdelyi}.

Changing the variable $\tau \rightarrow st$ \ and the parameter $zs^{\mu
}\rightarrow \zeta $ yields

\begin{equation}
\int\limits_{0}^{\infty }dte^{-st}E_{\mu }(-\zeta t^{\mu })=\frac{s^{\mu -1}%
}{s^{\mu }+\zeta }.  \label{eqA1}
\end{equation}

By differentiating Eq.(\ref{eqA1}) $n$ times with respect to $\zeta $ we
obtain,

\begin{equation}
\int\limits_{0}^{\infty }dte^{-st}t^{\mu n}E_{\mu }^{(n)}(-\zeta t^{\mu })=%
\frac{n!\cdot s^{\mu -1}}{(s^{\mu }+\zeta )^{n+1}}.  \label{eqA2}
\end{equation}

It is easy to see that at $\zeta =\mu $ Eq.(\ref{eqA2}) goes into Eq.(\ref%
{eq24}).


\begin{thebibliography}{99}
\bibitem{Laskin1} N. Laskin, Fractional quantum mechanics and L\'evy path
integrals, Phys. Lett. A\textbf{268} (2000) 298-305.

\bibitem{Laskin2} N. Laskin, Fractional quantum mechanics, Phys. Rev. E%
\textbf{62} (2000) 3135--3145 (also available online:
http://arxiv.org/PS\_cache/arxiv/pdf/0811/0811.1769v1.pdf).

\bibitem{Laskin3} N. Laskin, Fractals and quantum mechanics, Chaos \textbf{10%
} (2000) 780-790.

\bibitem{Laskin4} N. Laskin, Fractional Schr\={o}dinger equation, Phys. Rev.
E\textbf{66} (2000) 3135--3145 (also available online:
http://arxiv.org/PS\_cache/quant-ph/pdf/0206/0206098v1.pdf).

\bibitem{Laskin5} N. Laskin, L\'{e}vy flights over quantum paths,
Communications in Nonlinear Science and Numerical Simulation \textbf{12}
(2007) 2-18 (also available online:
http://arxiv.org/PS\_cache/quant-ph/pdf/0504/0504106v1.pdf).

\bibitem{Laskin6} N. Laskin, Principles of Fractional Quantum mechanics, in
J. Klafter, S.C Lim, R. Metzler (Eds.), Fractional Dynamics, Recent Advances
(World Scientific, Singapore, 2011), pp.393-427.

\bibitem{Feynman1} R.P. Feynman, The Space-Time Formulation of
Nonrelativistic Quantum Mechanics, Rev. Mod. Phys. \textbf{20 }(1948)\textbf{%
\ }367--387.

\bibitem{Feynman2} R. P. Feynman and A.R. Hibbs, Quantum Mechanics and Path
Integrals\textit{,} McGraw-Hill, New York, 1965.

\bibitem{Laskin7} N. Laskin, Fractional Poisson process, Communications in
Nonlinear Science and Numerical Simulation, \textbf{8} (2003) 201-213.

\bibitem{Kuno} M. Kuno, D.P. Fromm, H. F. Hamann, A. Gallagher, and D.J.
Nesbitt, Nonexponential blinking \ kinetics of single CdSe quantum dots: A
universal power law behavior, J. Chem. Phys. \textbf{112} (2000) 3117-3120.

\bibitem{Sabatelli} L. Sabatelli et al, Waiting time distributions in
financial markets, The European Physical Journal B - Condensed Matter and
Complex Systems, \textbf{2}7 (2002) 273-275.

\bibitem{Willinger} W. Willinger, and V. Paxson, Where Mathematics Meets the
Internet,\ Notices of the American Mathematical Society, \textbf{45 }(1998)
961-970.

\bibitem{Zaslavsky} A. I. Saichev and G.M. Zaslavsky, Fractional kinetic
equations: solutions and applications, Chaos \textbf{7} (1997) 753-764.

\bibitem{Laskin8} N. Laskin, Some applications of the fractional Poisson
probability distribution, J Math Phys \textbf{50} (2009) 113513.

\bibitem{Laskin9} N. Laskin, New Polynomials and Numbers Associated with
Fractional Poisson Probability Distribution, ICNAAM 2010: International
Conference of Numerical Analysis and Applied Mathematics 2010, Rhodes
(Greece), AIP Conference Proceedings \textbf{1281}, (2010) 1152-1155. (also
available online: http://arxiv.org/PS\_cache/arxiv/pdf/1010/1010.2874v1.pdf).

\bibitem{Klauder1} J.R. Klauder, The Current State of Coherent States,
arXiv:quant-ph/0110108v1 17 Oct 2001.

\bibitem{Bell} E. T. Bell, Exponential Polynomials, The Annals of
Mathematics, Second Series, \textbf{35} (1934), 258-277.

\bibitem{Stirling} J. Stirling, Methodus Differentialis: Sive Tractatus de
Summatione et Interpolatione Serierum Infinitarum, Gul. Bowyer, London,
1730. (English translation by I. Tweddle, James Stirling's Methodus
Differentialis: An Annotated Translation of Stirling's Text Springer,
London, 2003).

\bibitem{Charalambides1} C. A. Charalambides, Enumerative combinatorics,
Ch.8, Chapman\&Hall/CRC, 2002.

\bibitem{Charalambides} C. A. Charalambides and J. Singh: A review of the
Stirling numbers, their generalizations and statistical applications,
Communications in Statistics - Theory and Methods, \textbf{17} (1988)
2533-2595.

\bibitem{Dobinski} G. Dobi\'{n}ski, Summirung der Reihe $\sum n^{m}/n!$, f%
\"{u}r $m=1,2,3,4,5,...$, Grunert Archiv Arch. f\"{u}r M. und Physik \textbf{%
61} (1877) 333-336.

\bibitem{Rota} Gian-Carlo Rota, The Number of Partitions of a Set, American
Mathematical Monthly, \textbf{71} (1964) 498--504.

\bibitem{Kubo} N. Asai, I. Kubo and H. Kuo, Multiplicative renormalization
and generating functions I, Taiwaneese Journal of Mathematics, \textbf{7}
(2003) 89-101.

\bibitem{Klauder} J.R. Klauder, E.C.G. Sudarshan, Fundamentals of Quantum
Optics, Benjamin, New York, 1968.

\bibitem{Glauber} R. J. Glauber, Quantum Theory of Optical Coherence.
Selected Papers and Lectures, Wiley-VCH, Weinheim 2007.

\bibitem{Wolf} L. Mandel and E. Wolf, Optical Coherence and Quantum Optics, 
\textit{\ }Cambridge University Press, 1995.

\bibitem{Chihara} T. S. Chihara, An Introduction to Orthogonal Polynomials
(Mathematics and Its Applications), Gordon and Breach, Science Publishers,
New York, NY, 1978.

\bibitem{Oldham} K.B. Oldham and J. Spanier,\textit{\ }The Fractional
Calculus\textit{,} Academic Press, New York, 1974.

\bibitem{Samko} S.G. Samko, A.A. Kilbas, and O.I. Marichev, Fractional
Integrals and Derivatives and Their Applications, Gordon and Breach Science
Publishers, Langhorne, PA, 1993.

\bibitem{Miller} K.S. Miller and B. Ross, An Introduction to the Fractional
Calculus and Fractional Differential Equations, J. Wiley \& Sons, New York,
1993.

\bibitem{ML} G. Mittag-Leffler, Sur la repri\'esentation analytique d'une
branche uniforme d'une fonction monog\`ene, Acta Mathematics, \textbf{29}
(1905) 101-182.

\bibitem{Erdelyi} A. Erd\'elyi, Ed., Higher Transcendental Functions\textit{,%
} Vol.3, pp. 206-227 (Chapter 18 Miscellaneous Functions) McGraw-Hill, New
York, 1955.

\bibitem{Schrodinger} E. Schr\"{o}dinger, Der stetige \"{U}bergang von der
Mikro-zur Makromechanik, Die Naturwissenschaften \textbf{14 }(1926) 664-666.

\bibitem{Solomon} J.-M. Sixdeniers, K. A. Penson, and A. I. Solomon, J.
Phys. \textbf{A32}\ (1999) 7543-7564.

\bibitem{Katriel} J. Katriel, Bell numbers and coherent states, Phys. Lett. 
\textbf{A 273} (2000) 159-161.

\bibitem{Abramowitz3} Abramowitz, M. and Stegun, I. A. (Eds.). "Beta
Function" \S 6.2 in Handbook of Mathematical Functions with Formulas,
Graphs, and Mathematical Tables, 9th printing, p. 258, New York, Dover, 1972.

\bibitem{Abramowitz} M. Abramowitz and I.A. Stegun, (Eds.). Stirling Numbers
of the Second Kind. \S 24.1.4 in Handbook of Mathematical Functions with
Formulas, Graphs, and Mathematical Tables, 9th printing, pp. 824-825, New
York, Dover, 1972.

\bibitem{Abramowitz2} M. Abramowitz and I.A. Stegun, (Eds.). Stirling
Numbers of the First Kind. \S 24.1.3 in Handbook of Mathematical Functions
with Formulas, Graphs, and Mathematical Tables, 9th printing, p. 824, New
York, Dover, 1972.

\bibitem{Abramowitz1} M. Abramowitz and I.A. Stegun, (Eds.). Bernoulli and
Euler Polynomials and the Euler-Maclarin Formula \S 23.1 in Handbook of
Mathematical Functions with Formulas, Graphs, and Mathematical Tables, 9th
printing, pp. 804, New York, Dover, 1972.

\bibitem{Schlafli} L. Schl\"{a}fli \textquotedblleft On a generalization
given by Laplace of Lagrange's Theorem\textquotedblright , Quarterly Journal
of Pure and Applied Mathematics, \textbf{2} (1858) 24-31.

\bibitem{Tanny} S.M. Tanny, On some numbers related to the Bell numbers,
Canad. Math. Bull. \textbf{17} (1975), 733-738.

\bibitem{Mathworld} E.W. Weisstein, Poisson Distribution, From \textit{%
MathWorld} - A Wolfram Web Resource,

http://mathworld.wolfram.com/PoissonDistribution.html.
\end{thebibliography}
\end{document}